\documentclass[reviewcopy]{elsarticle}

\usepackage[reviewcopy]{adndt}
\usepackage{longtable}
\usepackage{tipa}
\usepackage{amsmath}
\usepackage{amssymb}

\usepackage{fdsymbol}

\biboptions{square,sort&compress}
\bibpunct[]{[}{]}{,}{n}{}{;}
\begin{document}

\begin{frontmatter}

\journal{Atomic Data and Nuclear Data Tables}

\title{\normalfont\textsc{Systematic study of fusion barriers with energy dependent barrier radius}}

  \author[One]{Yeruoxi Chen}

  \author[One,Two]{Hong Yao}

  \author[One,Three]{Min Liu}

  \author[One,Three]{Junlong Tian}

  \author[Four]{Peiwei Wen}

  \author[One,Three]{Ning Wang\corref{cor1}}
  \ead{E-mail: wangning@gxnu.edu.cn}

  \cortext[cor1]{Corresponding author.}

  \address[One]{Department of Physics, Guangxi Normal University, Guilin 541004, PR China}

  \address[Two]{School of Physics, Beihang University, Beijing 102206, PR China}

  \address[Three]{Guangxi Key Laboratory of Nuclear Physics and Technology, Guilin 541004, PR China}

  \address[Four]{China Institute of Atomic Energy, Beijing 102413, PR China}

\date{16.12.2002}

\begin{abstract}

Considering energy dependence of the barrier radius in heavy-ion fusion reactions, a modified Siwek-Wilczy\'{n}ski (MSW) fusion cross section formula is proposed. With the MSW formula, the fusion barrier parameters for 367 reaction systems are systematically extracted, based on 443 datasets of measured cross sections. We find that the fusion excitation functions for about $60\%$ reaction systems can be better described by introducing the energy dependence of the barrier radius which is due to the dynamical effects at energies near and below the barrier. Considering both the influence of the geometry radii and that of the reduced de Broglie wavelength of the colliding nuclei, the barrier heights are well reproduced with only one model parameter. The extracted barrier radius parameters linearly decrease with the effective fissility parameter, and the width of the barrier distribution relates to the barrier height, as well as the reduced de Broglie wavelength at energies around the Coulomb barrier.

{\vspace{0cm}} {\bf Key Words: }
fusion cross section, Coulomb barrier, barrier parameters, de Broglie wavelength
\end{abstract}

\end{frontmatter}

\newpage
\tableofcontents
\listofDtables
\listofDfigures
\vskip5pc

\section{Introduction}

The problem of overcoming a potential barrier is of importance not only in nuclear physics, but also in many other fields of the nature sciences. Knowledge of the nucleus-nucleus interaction potential is
an essential ingredient in the analysis of elastic and inelastic
scattering, as well as of fusion reactions between nuclei. The information concerning the potential barrier is of crucial importance for the synthesis of super-heavy nuclei and heavy-ion fusion at deep sub-barrier energies which has attracted a great deal of attention in recent years \cite{RevModPhys.72.733,PhysRevLett.104.142502,PhysRevC.78.034610,adamian2000dynamical,PhysRevC.84.061601,PhysRevC.84.064614,PhysRevC.62.044610,dasgupta1998measuring,PhysRevC.92.064604,PhysRevC.103.054601}. Up to now, the fusion cross sections for more than a thousand of reaction systems have been measured in the past several decades. A systematic study of the fusion barriers based on these data is therefore interesting and necessary.

Classically a particle can only overcome a potential barrier when its total energy exceeds the barrier height. In the classical description of fusion
excitation functions, the fusion cross section $\sigma_{\rm fus} (E)$
at a center-of-mass incident energy $E$ is given by
\begin{equation}
	\sigma_{\rm fus}(E) = \pi R_B^2(1-V_B/E),
\end{equation}
where $R_B$ is the barrier radius and $V_B$ is
the barrier height. The barrier parameters are obtained from the data
by fitting a straight line through a plot of $\sigma_{\rm fus}$  vs
$1/E$. The slope and intercept of this line with the
$1/E$ axis lead to the barrier radius and
height, respectively. At energies below the barrier, the particle may tunnel through the potential barrier, as a consequence of quantum mechanics. This tunnelling effect was first recognized in the 1920's and the $\alpha$-decay of nuclei was explained as a tunnelling effect \cite{gamow1928quantentheorie}. In the 1970's, the fusion cross sections are analytically described by the well known Wong formula \cite{PhysRevLett.31.766}, based on the assumption of a parabolic barrier together with the barrier penetration concept,
\begin{equation}
	\sigma_{\rm fus}(E) = \frac{\hbar\omega}{2E} R_B^2 \ln \{1+\exp[2\pi (E-V_B)/\hbar\omega]\} ,
\end{equation}
where, $\hbar \omega$ is the s-wave barrier curvature. The energy dependence of the barrier curvature is introduced in Ref. \cite{Denisov18} for a better description of the fusion cross sections at deeply sub-barrier energies. For relatively large values of $E$, the result of Wong formula reduces to the classical formula Eq.(1).

For the sub-barrier fusion reactions leading to heavy compound nuclei, an important observation is that the measured fusion cross sections exhibit strong enhancements compared to estimations using a simple one-dimensional barrier penetration model \cite{dasgupta1998measuring}. These enhancements have been accounted for in terms of strong couplings between the relative motion of colliding nuclei and the intrinsic degrees of freedom, such as the collective vibrations of nuclei and nucleon transfer in the neck region. To consider the coupling effects, Stelson introduced a distribution of barrier heights $D(B)$ in the calculation of the fusion excitation function around 1990 \cite{stelson1988neutron,PhysRevC.41.1584},
\begin{eqnarray}
\sigma_{\rm fus} (E)=\int D(B) \sigma
_{\rm fus}^{\rm (1)}(E,B)dB,
\end{eqnarray}
with $\int D(B) dB=1$.
Here, $\sigma_{\rm fus}^{\rm (1)}(E,B)$ is the fusion cross section based on a single barrier with a height $B$.

To describe $D(B)$, a single-Gaussian distribution of barrier heights predicted from different orientations of colliding nuclei undergoing slow deviations from sphericity is used by Siwek-Wilczy\'{n}ska and Wilczy\'{n}ski (SW). Together with Eq.(1) for describing $\sigma_{\rm fus}^{\rm (1)}(E,B)$, the SW formula was proposed \cite{PhysRevC.69.024611}. With the SW formula, the heavy-ion fusion cross sections for 29 systems, from $^{16}$O+$^{18}$O to $^{64}$Ni+$^{124}$Sn, at extreme sub-barrier
energies have been analyzed \cite{Jiang18}. Very recently, Wen et al. applied the SW formula to systematically extract the barrier information from the experimental fusion excitation functions, and found that the SW formula behaves much better for the barrier fitting than the Wong formula \cite{PhysRevC.105.034606}. In addition to the single-Gaussian function for describing $D(B)$, two-Gaussian function \cite{liu2006applications,wang2009heavy}, asymmetric Gaussian function \cite{PhysRevC.65.014607,WANG2017281} and as well as multi-Gaussian function \cite{wang2007systematic,PhysRevC.105.064601} are also frequently used. The experimental barrier height distribution $D(E)$ can be extracted from a precise experiment of the fusion excitation function via the second
derivative \cite{ROWLEY199125,TIMMERS1998421}:
\begin{equation}\label{eq.sdiff}
	D(E)=\frac{1}{\pi R_{B}^{2}} \frac{d^{2}\left(E \sigma_{\text {fus}}\right)}{d E^{2}}.
\end{equation}

The validity of all these analyses mentioned above requires that all $l$ waves contributing to the fusion cross section having the same barrier radius $R_B$, a condition which is probably not fulfilled for most reactions \cite{PhysRevC.17.126}. The energy dependence of the nucleus-nucleus potential was clearly observed from some microscopic dynamics calculations, such as the simulations based on the time-dependent Hartree-Fock (TDHF) theory \cite{PhysRevC.78.024610, PhysRevC.105.024328} and the improved quantum molecular dynamic (ImQMD) model \cite{PhysRevC.81.044602,PhysRevC.89.064601}, due to the strong dynamical effects in fusion process. For fusion reactions with deformed nuclei, the orientation of the colliding nuclei significantly influences not only the barrier height but also the barrier radius. For tip-tip configuration in fusion reaction induced by prolate nuclei, one obtains a larger barrier radius, comparing with that for side-side configuration \cite{PhysRevC.74.021601,PhysRevC.105.024328}. It is therefore necessary to investigate the influence of energy and orientation dependence of the barrier radius on the fusion cross sections.

The purpose of the present work is to systematically extract the fusion barriers based on the SW formula together with the energy dependence of the barrier radius being considered. The structure of this paper is as follows: In Sec. 2, energy dependence of the barrier radius for $^{16}$O+$^{208}$Pb and $^{34}$S+$^{168}$Er will be investigated. In Sec. 3, a modified SW formula will be proposed and the model accuracy for describing some fusion excitation functions will be tested. In Sec. 4, the information concerning fusion barriers extracted from 443 datasets of experimental data for 367 different projectile-target combinations, will be presented and the systematics of the fusion barrier will also be analyzed. Finally, a summary will be given in Sec. 5.

\section{Energy dependence of barrier radius}

 \begin{figure}
    \centering
    \includegraphics[width=0.7\textwidth]{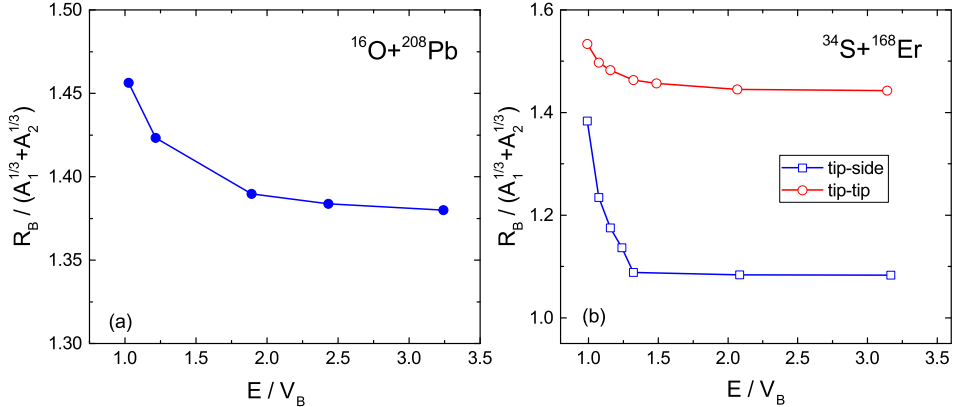}
    \caption{ Barrier radius as a function of $E/V_B$ for $^{16}$O+$^{208}$Pb (a) and $^{34}$S+$^{168}$Er (b) from the TDHF calculations.  $A_1$ and $A_2$ denote the mass number of projectile and target nuclei, respectively. $V_B$ in horizontal axis are taken from a linear fit of the measured $\sigma_{\rm fus}$ \cite{PhysRevC.60.044608,PhysRevC.62.024607} vs $1/E$ in the region larger than 100 mb.}
  \end{figure}

To investigate the energy dependence of barrier radius, we firstly use the time dependent Hartree-Fock (TDHF) theory for simulating the fusion reactions $^{16}$O+$^{208}$Pb and $^{34}$S+$^{168}$Er. The nucleus-nucleus interaction potential is extracted by using the density-constrained TDHF approach \cite{PhysRevC.74.021601,PhysRevC.80.041601}. In $^{34}$S+$^{168}$Er, the tip-tip and tip-side orientations for the deformed reaction partners are taken into account. The Skyrme SLy4d interactions [26] are used by static HF and TDHF dynamic evolution, in which the numerical boxes are chosen as $30\times30\times30$ fm$^3$ and $30\times 30\times 50$ fm$^3$, respectively. The time propagation is carried out using a Talyor-seriers expansion up to the sixth order of the unitary mean-field propagator with a time step of 0.2 fm/c, and the initial distance of two nuclei is set to 20 fm.

In Fig. 1, we show the calculated barrier radii at different incident energies for $^{16}$O+$^{208}$Pb and $^{34}$S+$^{168}$Er, which are scaled by $A_1^{1/3}+A_2^{1/3}$ to see the radius parameter.  One can see that for both reactions at energies around the barrier height $V_B$, the barrier radius decrease evidently with incident energy. Especially for $^{34}$S+$^{168}$Er at tip-side orientation, the radius parameter falls sharply with energy, from 1.4 fm at $E\approx V_B$ down to 1.1 fm at $E\approx 1.3 V_B$. At energies much higher than the barrier height, the barrier radius does not change too much. The energy dependence of the barrier radius is due to the dynamical effects in fusion process. At energies around the barrier height, the fusion process is relatively slow and the reaction partners have enough time to readjust nuclear density distributions of the reaction system. The dynamical deformation of the densities and neutron transfer in the neck region can result in the enlargement of the barrier radius and the reduction of barrier height correspondingly.

In addition to the TDHF calculations, the barrier radius is also analyzed based on the measured fusion excitation function. According to Eq.(1), the barrier radius can be expressed as,
\begin{equation}
	R_B(E) = \left [\frac{\sigma_{\rm fus}}{\pi(1-V_B/E)} \right ]^{1/2}.
\end{equation}
In Fig. 2, we show the extracted barrier radius for $^{16}$O+$^{208}$Pb and $^{34}$S+$^{168}$Er, based on the measured fusion excitation function $\sigma_{\rm fus}$ \cite{PhysRevC.60.044608,PhysRevC.62.024607}. At energies above the fusion barrier, the barrier radius does not change significantly with incident energy. At sub-barrier energies, the enhancement of the barrier radius can be clearly observed. In Ref. \cite{PhysRevC.91.044617}, a generalized Wong formula is proposed by Rowley and Hagino through considering energy dependence of the barrier parameters. The trend of the energy dependence for barrier radius observed in Fig. 2 is generally in agreement with those from the TDHF calculations and the generalized Wong formula.

  \begin{figure}
    \centering
    \includegraphics[width=0.7\textwidth]{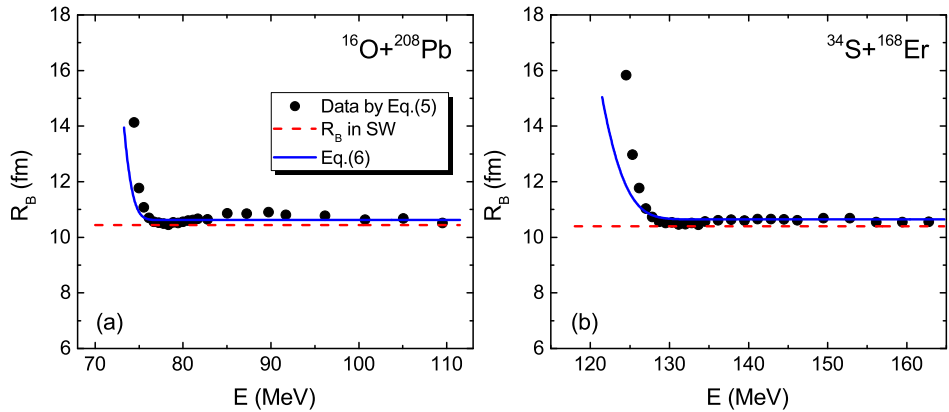}
    \caption{Barrier radius for for $^{16}$O+$^{208}$Pb (a) and $^{34}$S+$^{168}$Er (b) from the measured fusion excitation functions. The circles denote the extracted results from the data by using Eq.(5). The dashed lines denote the result of SW formula. The solid curves denote the results of Eq.(6).}
  \end{figure}

Both the TDHF calculations and the data analysis imply that the assumption used in the traditional formula, i.e., all $l$ waves contributing to the fusion cross section have the same barrier radius $R_B$, is generally valid at energies above the barrier height. However, at energies around the barrier, the enhancement of barrier radius due to dynamical effects should be considered for a better description of the fusion excitation functions. To consider the influence of the dynamical effects on barrier radius, we empirically introduce a correction term to the traditional barrier radius $R_0$,
\begin{equation}\label{eq.rbE}
	R_{B} (X) = R_{0} + \Delta R \exp (-X) \operatorname{erfc} \left(X\right).
\end{equation}
The definition of $X$ and the determination of the correction factor $\Delta R$ will be discussed later. The solid curves in Fig. 2 denote the results according to Eq.(6). Comparing with the results of SW formula, the trend of energy dependence for the barrier radius can be much better described by using Eq.(6), especially at sub-barrier energies.

\section{Modified Siwek-Wilczy\'{n}ski formula and some tests}

    \begin{figure}
     \centering
        \includegraphics[width=0.9\textwidth]{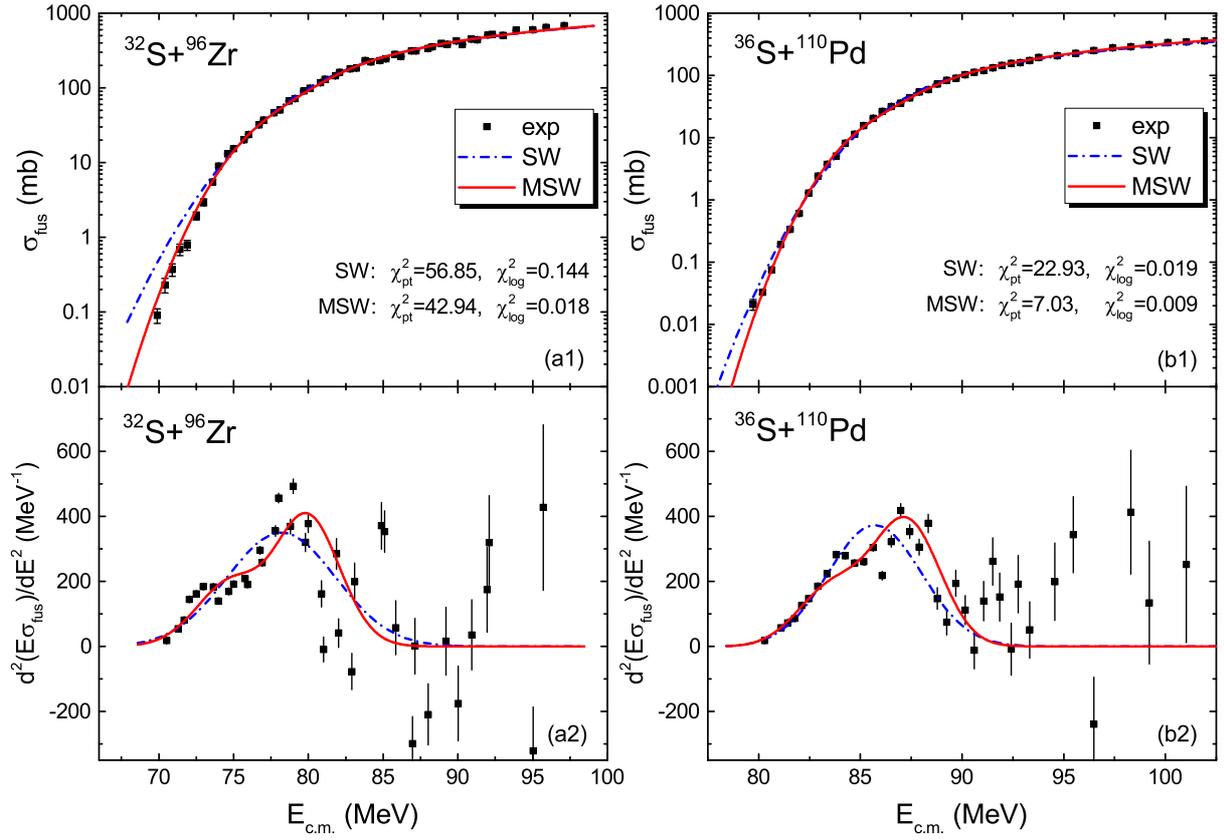}
        \caption{(color online) Fusion excitation functions and fusion barrier distributions for $^{32}$S+$^{96}$Zr and $^{36}$S+$^{110}$Pd. The squares in (a1) and (b1) denote the experimental data taken from \cite{PhysRevC.82.054609,PhysRevC.52.R1727} for $^{32}$S+$^{96}$Zr and $^{36}$S+$^{110}$Pd, respectively. The squares in (a2) and (b2) denote the extracted barrier distribution according to Eq.(4). The dash-dotted curve and the solid curve denote the results of SW formula and those of MSW formula, respectively.}
    \end{figure}

Considering the energy dependence of the barrier radius given by Eq.(6), we propose a modified Siwek-Wilczy\'{n}ski (MSW) formula for describing the fusion excitation function
%\begin{equation}\label{eq.SW}
%	\sigma_{\rm {fus}}(E)=\pi R_B^{2}(X) \frac{ W}{2E} \left[X \operatorname{erfc}(-X)+\frac{1}{\sqrt{\pi}}\exp\left(-X^{2}\right) \right ],
%\end{equation}
%where $X = (E- V_B)/W$. $V_B$ and $W$ denote the centroid and the width of the Gaussian function, respectively.

\begin{equation}\label{eq.MSW}
	\sigma_{\rm {fus}}(E)=\pi R_B^{2}(X) \frac{W}{\sqrt{2}E}
	[X \operatorname{erfc}(-X)+\frac{1}{\sqrt{\pi}}\exp(-X^{2}) ],
\end{equation}
where $X = \frac{E-V_B}{\sqrt{2}W}$. $V_B$ and $W$ denote the centroid and the standard deviation of the Gaussian function, respectively. Together with the traditional barrier radius $R_0$ and the correction factor $\Delta R$ in Eq.(6), there are a total of four barrier parameters in the MSW formula. If $\Delta R=0$, the result of Eq.(7) reduces to the standard SW formula \cite{PhysRevC.69.024611}. For a certain fusion reaction, the four parameters in the MSW formula can be determined by fitting the measured fusion excitation function. The popular Minuit minimization program \cite{james1975minuit} is usually applied to determine the fitting parameters by searching the global minimum in the hypersurface of the $\chi^2$
function. The $\chi^2$ per energy point is expressed as
\begin{equation}\label{eq.x2pt}
	\chi^{2}_{pt}=\frac{1}{N} \sum_{i=1}^{N}\left[\frac{\sigma_{\mathrm{th}}\left(E_{i}\right)-\sigma_{\exp }\left(E_{i}\right)}{\delta \sigma_{\exp }\left(E_{i}\right)}\right]^{2},
\end{equation}
in which the uncertainty of fusion cross section is involved in the fitting process. In addition to $\chi^2_{pt}$, the mean-square deviation between the measured fusion cross sections and model predictions is also frequently used to determine the best-fit model parameters \cite{WANG2017281,wang2007systematic}. Here, the mean-square deviation in logarithmic scale is defined as,
\begin{equation}\label{eq.x2log}
	\chi_{\log}^{2}=\frac{1}{N} \sum_{i=1}^{N}\left[\log \left(\sigma_{\mathrm{th}}\left(E_{i}\right)\right)-\log \left(\sigma_{\exp }\left(E_{i}\right)\right)\right]^{2}.
\end{equation}
$\chi_{\log }^{2}$ is more effective to check the trend of fusion cross sections at sub-barrier energies. In this work, we  combine these two quantities and use $\bar \chi =(\chi^{2}_{pt}+\chi^{2}_{\log })^{1/2}$ to search for the best-fit parameters.

   \begin{figure}
     \centering
        \includegraphics[width=0.9\textwidth]{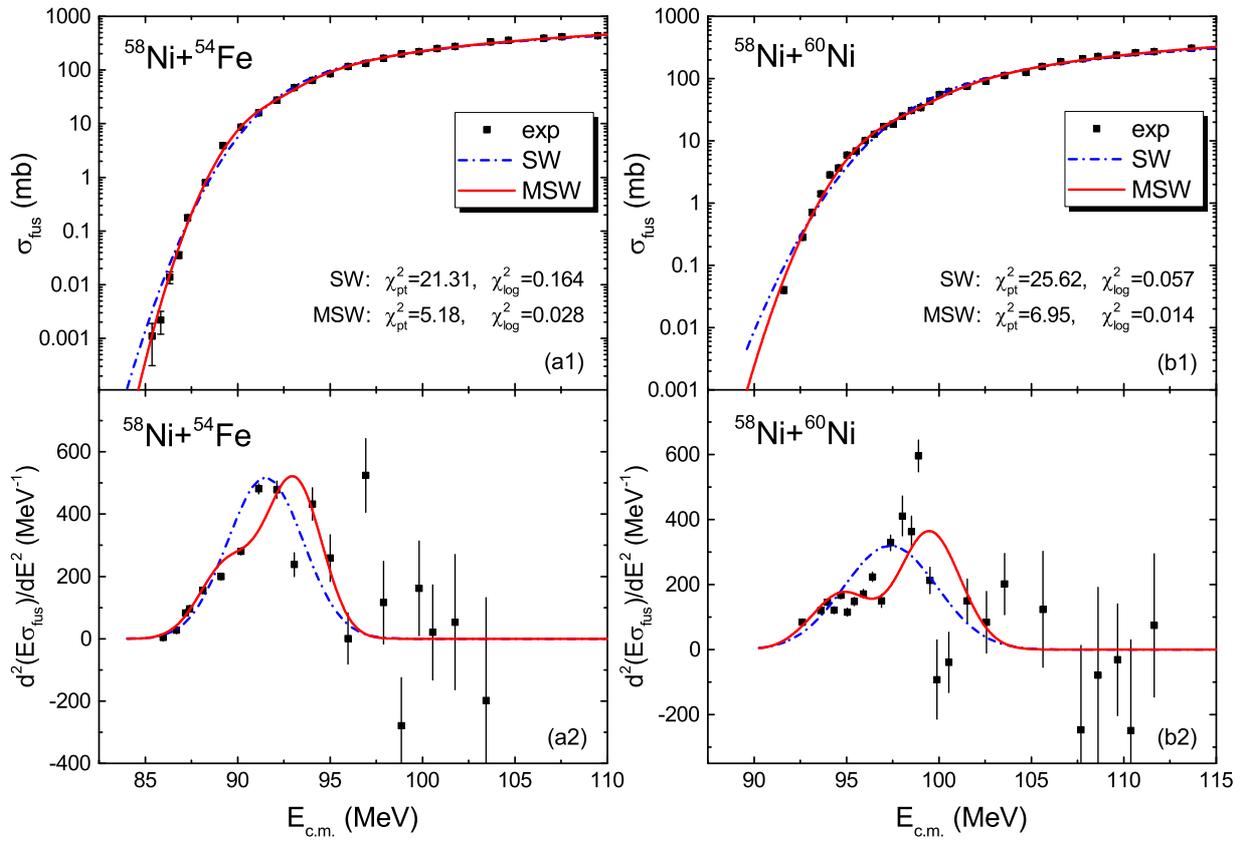}
        \caption{(color online) The same as Fig. 3, but for $^{58}$Ni+$^{54}$Fe and $^{58}$Ni+$^{60}$Ni. The experimental data are taken from \cite{PhysRevC.82.014614,PhysRevLett.74.864}.}
    \end{figure}

Figure 3 and Fig. 4 show the fusion excitation functions and barrier distributions for $^{32}$S+$^{96}$Zr, $^{36}$S+$^{110}$Pd, $^{58}$Ni+$^{54}$Fe and $^{58}$Ni+$^{60}$Ni reactions. We note that introducing the energy dependence of barrier radius, the experimental data can be much better reproduced, especially for the fusion cross sections at deep sub-barrier energies. With the MSW formula, both $\chi^{2}_{pt}$ and $\chi^{2}_{\log }$ are significantly smaller than those with the SW formula. In addition, the barrier distributions are also studied to check the details in reproducing the fusion excitation functions. The distributions are extracted from the experimental excitation functions using the point-difference approximation \cite{TIMMERS1998421} according to Eq.(4),
\begin{equation}
	\frac{d^{2}\left(E \sigma_{\text {fus}}\right)}{d E^{2}}\approx \frac{2E\sigma_{\rm fus}(E)-E\sigma_{\rm fus}(E+\Delta E)-E\sigma_{\rm fus}(E-\Delta E)}{(\Delta E)^2},
\end{equation}
with an energy step $\Delta E=2.5$ MeV. From (a2) and (b2) in Fig. 3 and Fig. 4, one can see that with energy dependence of $R_B$, the left shoulders in the barrier distributions for these four systems can be evidently observed in the MSW calculations, although the single-Gaussian function is adopted in Eq.(3).

\section{Extracted barrier parameters }

   \begin{figure}
     \centering
        \includegraphics[width=0.85\textwidth]{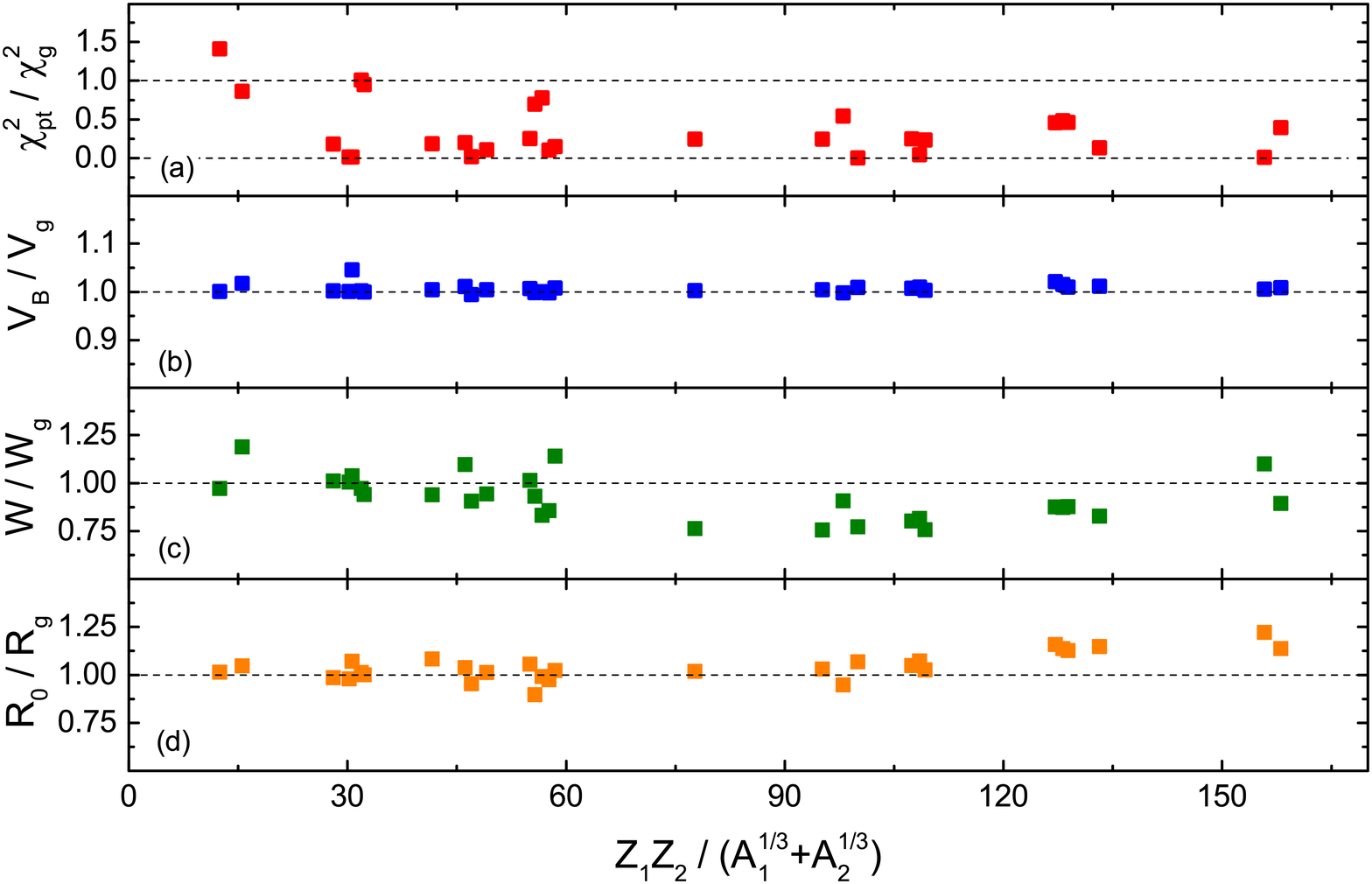}
        \caption{(color online) Ratios between the extracted barrier parameters in this work and those in Ref. \cite{Jiang18}.  $\chi^{2}_{g}$, $V_g$, $W_g$ and $R_g$ denote the obtained $\chi^{2}_{pt}$, the barrier height, the standard deviation and the barrier radius in Ref. \cite{Jiang18}, respectively.}
    \end{figure}

Up to now, a large number of fusion excitation functions have been measured in the past several decades. Most of data for the fusion cross sections including fusion-fission and evaporation residue obtained from the tables or the graphes of the corresponding publications, are collected in the NRV website \cite{nrvweb}. In this work, we use a similar procedure as adopted in Ref. \cite{PhysRevC.105.034606} to select the experimental fusion data. In addition to the data in the NRV website, some fusion excitation functions measured in very recent years are also collected in this work. One usually defines the fusion cross section $\sigma_{\rm fus}$ as a sum of evaporation residue cross section $\sigma_{\rm EvR}$ and fission cross section $\sigma_{\rm FF}$. For light and intermediate mass systems, it is thought that $\sigma_{\rm fus}\simeq \sigma_{\rm EvR}$, since the fission barrier of the compound nuclei is high enough and the fission cross sections could be negligible. For heavy systems, e.g. the reactions leading to lanthanides or heavier nuclei, the contribution of fission cannot be ignored, the fission cross sections need to be included in $\sigma_{\rm fus}$. For fusion reactions leading to actinides, the evaporation residue cross sections are relatively small and the fission cross sections play a dominant role in the extraction of the fusion barrier. For fusion reactions leading to super-heavy nuclei, the evaporation residues become negligible and the quasi-fission cross sections are dominant in the total capture cross sections with which the Coulomb barrier can be extracted. For some systems with the same projectile-target combination, the data from different experimental groups are slightly different and the fusion barrier is separately analyzed and presented in this work.

Firstly, we analyze the 29 fusion reaction systems mentioned in Ref. \cite{Jiang18}, where Jiang et al. systematically analyzed the fusion cross sections for the 29 systems by using the SW formula. In Fig. 5, we compare the results from the SW formula and those from the MSW formula proposed in this work. From Fig. 5(a), one notes that for most of reactions, the $\chi^{2}_{pt}$ values with the MSW formula are much smaller than those with SW formula, since one more parameter $\Delta R$ is involved. For the lightest system $^{16}$O+$^{18}$O, the obtained barrier parameters from the two formulas are very close to each other, although the obtained $\chi^{2}_{pt}$ in this work is larger than that in Ref. \cite{Jiang18}. The obtained barrier heights from the two formulas are in good agreement with each other. The discrepancies in $W$ and $R_0$ are within $25\%$, which indicates that the introduction of the correction factor $\Delta R$ in the MSW formula influences $W$ and $R_0$ relatively larger than $V_B$.

Then, we systematically analyze a total of 443 datasets of measured fusion (and/or fission) cross sections for 367 different projectile-target combinations by using the MSW formula. The values of $\bar\chi$ corresponding to the best-fit parameters for all considered systems are simultaneously obtained. In Fig. 6, we show the distribution for the relative deviation of $\bar\chi$ between the results of MSW and those of SW, i.e., $(\bar\chi^{\rm SW}-\bar\chi^{\rm MSW})/\bar\chi^{\rm SW}$. We find that for 173 datasets, the relative deviation $\Delta \bar\chi/\bar\chi$ is smaller than $0.1\%$. For 122 datesets, the improvement is larger than $10\%$ and for 148 datesets the values of $\Delta \bar\chi/\bar\chi$ are located in the region of $0.1\%-10\%$. It indicates that the measured fusion excitation functions for about $60\%$ reactions can be better reproduced by introducing the energy dependence of barrier radius into the SW formula. The extracted barrier parameters and the references for these systems are listed in Table A.

   \begin{figure}
     \centering
        \includegraphics[width=0.7\textwidth]{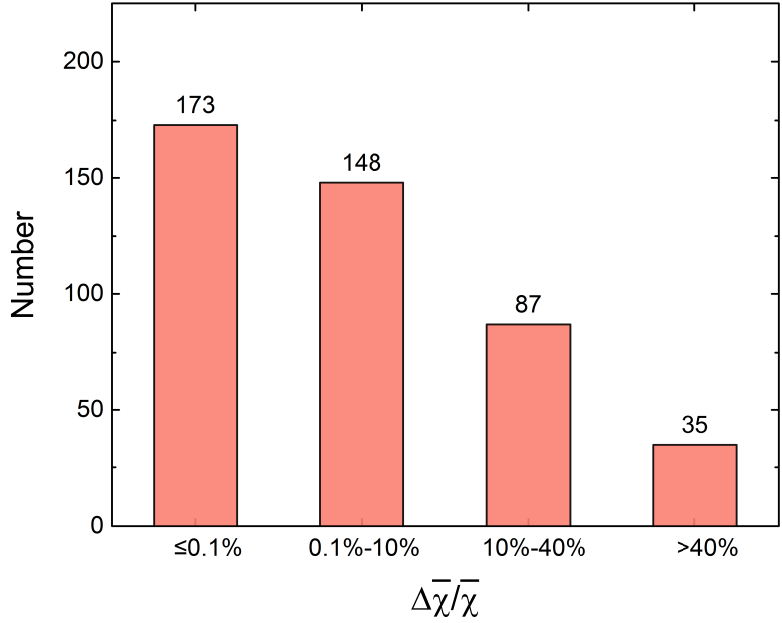}
        \caption{(color online) Distribution of the relative improvement in $\bar \chi$ by using MSW comparing with SW. }
    \end{figure}

For studying the fusion of heavy nuclei, one usually introduces a parameter to describe the fissility of the reaction system.
The effective fissility parameter is defined as ,
\begin{equation}\label{eq.Xeff}
	x_{\rm{eff}}=\frac{(Z^2/A)_{\rm eff}}{(Z^2/A)_{\rm thr}},
\end{equation}
with the effective fissility
\begin{equation}
	(Z^2/A)_{\rm eff}=\frac{4 Z_{1} Z_{2}}{A_{1}^{1 / 3} A_{2}^{1 / 3}({A}_{1}^{1/3}+A_{2}^{1/3})}
\end{equation}
and the threshold \cite{SWIATECKI1982275} for the effective fissility $(Z^2/A)_{\rm thr}\approx 33$, beyond which an extra push is needed to achieve fusion. $Z_1$ and $Z_2$ in Eq.(12) denote the charge numbers of the projectile and target nuclei, respectively.

Based on the extracted barrier height $V_B$, the radius of the corresponding Coulomb potential $R_{\rm Coul} = Z_{1}Z_{2}e^2/V_{B}$ is systematically analyzed. In Fig. 7(a), we show the extracted radius parameter $R_{\rm Coul}$. The decreasing trend of the radius parameter $R_{\rm Coul}/(A_1^{1/3}+A_2^{1/3})$ with the effective fissility parameter $x_{\rm eff}$ can be evidently observed. To understand the physics behind, we also show in Fig. 7(b) the reduced de Broglie wavelength $\middlebar{\lambda}_B=\hbar/\sqrt{2\mu V_B}$ of the colliding nuclei at an incident energy of $E =V_B$. $\mu$ is the reduced mass of the reaction system. One can see that $\middlebar{\lambda}_B$ approaches to zero with the increase of $x_{\rm eff}$, which indicates that the influence of de Broglie wavelength is negligible for heavy fusion systems. It is known that the capture cross section  $\sigma_{\rm cap}\propto \pi (R+ \middlebar{\lambda} )^2$ considering the wave properties of incident particles. For  heavy fusion system  $\middlebar{\lambda}$ is very small and consequently one obtains the traditional geometry cross section $\propto \pi R^2$. For very light fusion systems (with smaller values of $x_{\rm eff}$) and thermal neutron induced capture cross sections, the contribution of $\middlebar{\lambda}$ needs to be considered.

   \begin{figure}
     \centering
        \includegraphics[width=0.65\textwidth]{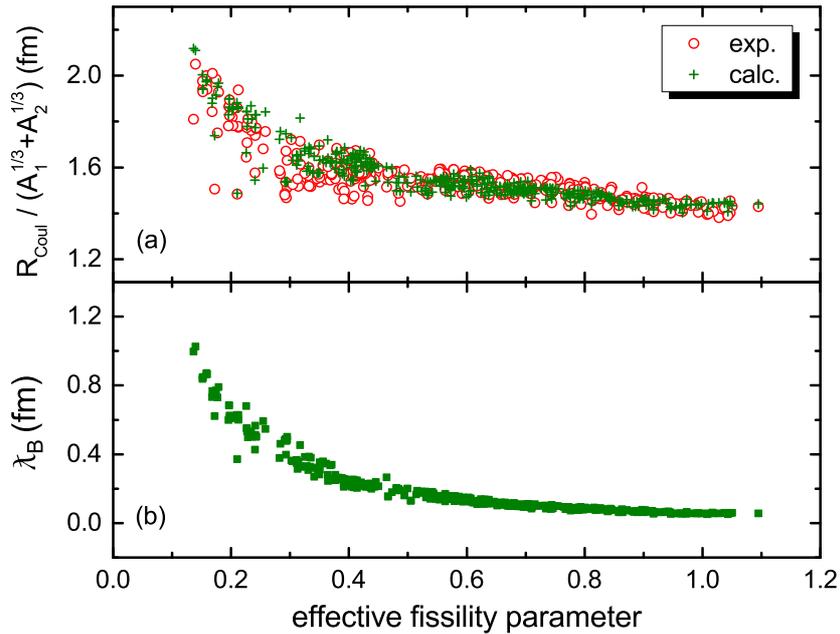}
        \caption{(color online) (a) Extracted radius parameter for the Coulomb potential as a function of the effective fissility parameter $x_{\rm{eff}}$. The open circles and the crosses denote the data based on the extracted barrier heights and the calculated results by using Eq.(13), respectively. (b) Reduced de Broglie wavelength $\middlebar{\lambda}_B$  of the colliding nuclei at an incident energy of $E =V_B$.}
    \end{figure}

To describe the radius of the Coulomb potential $R_{\rm Coul}$, we consider both the influence of the geometry radii of nuclei and that of the wave properties of particles. We write $R_{\rm Coul}$ as a sum of the charge radii of projectile and target nuclei, a parameter $d=1.75$ fm which is related to the interaction range, and as well as the reduced de Broglie wavelength,
\begin{equation}
	R_{\rm Coul} =R_1^C+R_2^C +1.75 + \middlebar{\lambda}_B.
\end{equation}
The charge radius $R^C \simeq\sqrt{\frac{5}{3}}r_{\rm ch}$ of a nucleus neglecting its deformations is taken from the root-mean-square (rms) charge radius $r_{\rm ch}$ which can be measured with high precision \cite{PhysRevC.88.011301,Angeli13,Litao21}. In the calculations, the reduced de Broglie wavelength $\middlebar{\lambda}_B=\hbar/\sqrt{2\mu V_B}$ can be obtained by using an iterative procedure with an initial value of $V_B\approx Z_1 Z_2 e^2/(R_1^C+R_2^C +1.75)$. The calculated results of $R_{\rm Coul}$ are also shown in Fig. 7(a) for comparison. One sees that the extracted radius parameter can be well reproduced. It indicates that the decreasing trend of the barrier radius parameter could have a relationship with the de Broglie wavelength of the colliding nuclei. With only one parameter in Eq.(13), the extracted barrier heights can be well reproduced by using
\begin{equation}
 V_B=Z_1 Z_2 e^2/R_{\rm Coul},
\end{equation}
with an rms deviation of only 1.52 MeV for all considered reactions. The relative deviation $\Delta V_B=(V_B^{\rm exp} -V_B^{\rm th})/V_B^{\rm exp}$ between data and model predictions is also calculated, the corresponding rms error is $2.83\%$ with Eq.(14) for calculating $V_B^{\rm th}$, which is much smaller than the corresponding value of $4.29\%$ from the three-parameter WKJ formula in Ref. \cite{Swiateck05} and slightly smaller than that of $2.84\%$ from the two-parameter MCW formula in Ref. \cite{PhysRevC.105.034606}.

   \begin{figure}
     \centering
        \includegraphics[width=0.65\textwidth]{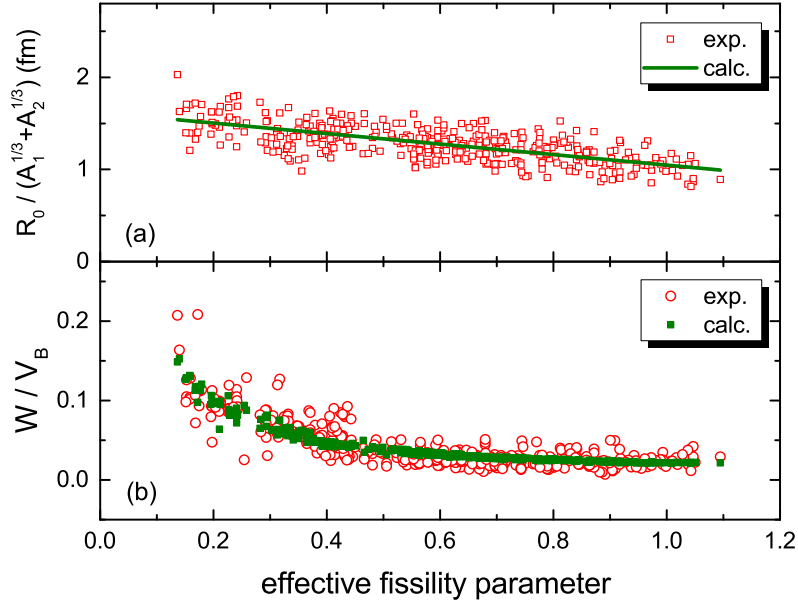}
        \caption{(color online) (a) Extracted barrier radius parameters as a function of the effective fissility parameter $x_{\rm{eff}}$. The line denotes a linear fit to the extracted results. (b) Standard deviation of the Gaussian function $W$ divided by the corresponding barrier height $V_B$ as a function of $x_{\rm eff}$. The circles denote the extracted results and the squares denote the predictions with Eq.(16).}
    \end{figure}

The systematics of the extracted barrier radius $R_0$ and that of the standard deviation of the Gaussian function $W$ are investigated simultaneously. In Fig. 8, we show the extracted barrier radius parameters $R_0/(A_1^{1/3}+A_2^{1/3})$ and the value $W/V_B$ as functions of the effective fissility parameter $x_{\rm{eff}}$. We note that the extracted barrier radius parameter linearly decreases with the effective fissility parameter, and the decreasing trend of $W/V_B$ is very similar to that of $\middlebar{\lambda}_B$ in Fig. 7(b). We therefore propose two formulas,
\begin{equation}
 R_0 = (1.62-0.57x_{\rm eff})(A_1^{1/3}+A_2^{1/3})
\end{equation}
 and
\begin{equation}
 W=(0.014+0.135\middlebar{\lambda}_B)V_B,
\end{equation}
for describing $R_0$ and $W$, respectively. The similar trends for the barrier radius and the distribution width are also observed in Ref. \cite{Jiang18}. In addition to the influence of wave properties of reaction partners, the systematic decreasing trend of the barrier radius $R_0$ could also be due to the influence of quasi-fission of reaction systems, since the quasi-fission cross sections are not involved in the present analysis. The contribution of quasi-fission to the total capture cross sections may increase with $x_{\rm{eff}}$. It is thought that the influence of quasi-fission becomes evident and an extra push is needed to achieve fusion for the systems with $x_{\rm{eff}}>1$ \cite{SWIATECKI1982275,BOCK1982334}. The systematics of the correction factor $\Delta R$ is unclear at the moment. The shell effects of reaction systems and the change of $Q$ value due to nucleon transfer should be further investigated to explore the systematics of $\Delta R$.

   \begin{figure}
     \centering
        \includegraphics[width=0.7\textwidth]{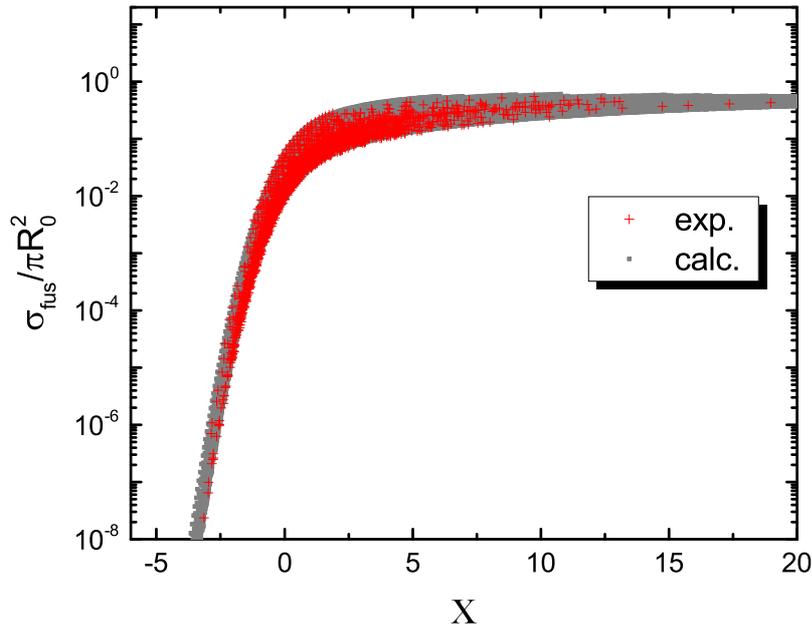}
        \caption{(color online) Measured cross sections for reactions with $\Delta R=0$ as a function of $X = \frac{E-V_B}{\sqrt{2}W}$. Here, the cross sections are scaled by $\pi R_0^2$. Eq.(14), Eq.(15) and Eq.(16) are used in the calculations of $V_B$, $R_0$ and $W$, respectively. Squares denote the model predictions from the SW formula. }
    \end{figure}

For fusion reactions with $\Delta R=0$ listed in Table A, we systematically analyze the experimental data. Simultaneously, the fusion cross sections for these systems are also calculated for comparison by using the SW formula with Eqs.$(14)-(16)$ for calculating $V_B$, $R_0$ and $W$. In Fig. 9, the measured cross sections scaled by $\pi R_0^2$ are shown as a function of $X$. We note that the cross sections for different reactions have a quite similar trend at sub-barrier and over-barrier energies. The experimental data can be reasonably well reproduced by the SW formula together with the proposed barrier parameter formulas.

\section{Summary}

A total of 443 datasets of measured fusion (and/or fission) excitation functions for 367 different projectile-target combinations, are systematically analyzed by using a new fusion cross section formula, in which the energy-dependent barrier radius $R_B$ is introduced into the Siwek-Wilczy\'{n}ski formula. We find that the fusion excitation functions for about $60\%$ reaction systems can be better described by considering the energy dependence of $R_B$. The energy dependence of barrier radius can also be clearly observed from the time-dependent Hartree-Fock (TDHF) calculations, which is due to the dynamical effects at energies around the Coulomb barrier. With the energy dependence of $R_B$, the barrier distributions based on the double-differentiation process  can be better reproduced for some systems, especially the left shoulder in the distribution. Considering both the influence of the geometry radii and that of the wave properties of the colliding nuclei, the barrier height $V_B$ can be well reproduced with only one model parameter. We also note that the extracted barrier radius parameters linearly decrease with the effective fissility parameter, from about $1.6$ fm for very light reaction systems to about $1.0$ fm for heavy systems in which quasi-fission could occur. It seems that the width of the barrier distribution relates to the barrier height, as well as the reduced de Broglie wavelength of the colliding nuclei at energies around the Coulomb barrier.

\ack
This work was supported by National Natural Science Foundation of China (Grant No. 12265006, U1867212), Guangxi Natural Science Foundation (Grant NO. 2023GXNSFDA026005, 2017GXNSFGA198001) and Middle-aged and Young Teachers' Basic Ability Promotion Project of Guangxi (CN) (Grant No. 2019KY0061). The authors would like to thank Huiming Jia and Li Ou for helpful discussions.

\newpage
\begin{appendix}

	\def\thesection{} % To get the appendix heading correct
	\section{Appendix}

\setlength{\LTleft}{0pt}
\setlength{\LTright}{0pt}
\setlength{\tabcolsep}{0.5\tabcolsep}
\renewcommand{\arraystretch}{1.0}
\footnotesize

\begin{longtable}{llllllllll}
\caption{ Extracted barrier parameters based on 443 datasets of measured cross sections together with the MSW formula. $V_{B}$ and $W$ denote the centroid and the standard deviation of the Gaussian function for barrier distribution, respectively. $R_{0}$ denotes the traditional barrier radius. $\Delta R$ denotes the correction factor for the barrier radius.  $\chi^2_{pt}$ denotes the $\chi^2$ per energy point, and $\chi_{\log }^{2}$ denotes the mean-square deviation between data and predictions in logarithmic scale. $\rm {EvR}$ and $\rm {FF}$ are cross sections for fusion-evaporation residues and fusion-fission, respectively. \\ }
Reaction~~~~~~~~~ & $V_{B}$(MeV)~~~~~~ & $W$(MeV) ~~~~ & $R_{0}$(fm)~~~~~~~ & $\Delta R$(fm)~~~~~~~ & $\chi^2_{pt}$~~~~~~~~~~~& $\chi_{\log}^{2} (\times 100)$~~~~~ & Type~~~~~~~~~ & Ref.\\
\hline\\
\endfirsthead
\caption[]{(continued)}
Reaction~~~~~~~~~ & $V_{B}$(MeV)~~~~~~ & $W$(MeV) ~~~~ & $R_{0}$(fm)~~~~~~~ & $\Delta R$(fm)~~~~~~~ & $\chi^2_{pt}$~~~~~~~~~~~ & $\chi_{\log}^{2} (\times 100)$~~~~~ & Type~~~~~~~~~ & Ref.\\
\hline\\
\endhead
$^{4}$He+$^{93}$Nb & 11.74 & 1.00 & 9.38 & 0.350 & 0.609 & 0.666 & EvR & \cite{PhysRevC.82.044608} \\
$^{4}$He+$^{154}$Sm & 15.29 & 0.39 & 9.77 & 3.744 & 0.393 & 0.398 & EvR & \cite{PhysRevC.31.1752} \\
$^{4}$He+$^{233}$U & 22.63 & 1.67 & 11.83 & 0.244 & 5.418 & 14.639 & FF & \cite{FREIESLEBEN1974503} \\
$^{4}$He+$^{235}$U & 23.28 & 2.06 & 10.49 & 0.003 & 0.041 & 0.042 & FF & \cite{PhysRev.115.1247} \\
$^{4}$He+$^{236}$U & 22.25 & 1.59 & 10.55 & 0 & 0.169 & 0.124 & FF & \cite{LIMKILDE1973504} \\
$^{4}$He+$^{238}$U & 22.90 & 1.67 & 12.57 & 0.242 & 4.341 & 9.856 & FF & \cite{FREIESLEBEN1974503} \\
$^{4}$He+$^{238}$U & 23.00 & 2.02 & 10.04 & 0.001 & 0.024 & 0.024 & FF & \cite{PhysRev.121.1415} \\
$^{4}$He+$^{237}$Np & 21.12 & 0.64 & 9.39 & 2.391 & 0.810 & 0.877 & EvR+FF & \cite{PhysRevC.7.1231} \\
$^{6}$He+$^{64}$Zn & 9.86 & 2.06 & 8.14 & 0 & 0.015 & 0.083 & EvR & \cite{fisichella2011halo} \\
$^{6}$He+$^{209}$Bi & 19.57 & 2.23 & 9.67 & 0 & 0.496 & 0.457 & EvR & \cite{PhysRevLett.81.4580} \\
$^{8}$He+$^{197}$Au & 19.61 & 1.95 & 11.41 & 0 & 0.196 & 0.331 & EvR & \cite{PhysRevLett.103.232701} \\
$^{6}$Li+$^{64}$Ni & 11.82 & 1.24 & 8.57 & 0 & 0.089 & 0.165 & EvR & \cite{PhysRevC.91.034615} \\
$^{6}$Li+$^{64}$Zn & 12.68 & 1.63 & 8.75 & 0.161 & 0.392 & 1.802 & EvR & \cite{PhysRevC.87.064614} \\
$^{6}$Li+$^{90}$Zr & 17.45 & 1.05 & 8.17 & 2.478 & 1.351 & 1.180 & EvR & \cite{PhysRevC.86.024607} \\
$^{6}$Li+$^{144}$Sm & 25.02 & 1.66 & 7.70 & 0.145 & 0.931 & 4.811 & EvR & \cite{PhysRevC.79.051601} \\
$^{6}$Li+$^{152}$Sm & 24.47 & 1.95 & 8.26 & 0 & 0.890 & 0.301 & EvR & \cite{RATH201214} \\
$^{6}$Li+$^{159}$Tb & 24.27 & 2.01 & 8.08 & 0.189 & 0.433 & 0.353 & EvR & \cite{PhysRevC.83.064606} \\
$^{6}$Li+$^{198}$Pt & 28.69 & 1.82 & 8.45 & 0.192 & 0.426 & 2.926 & EvR & \cite{PhysRevLett.103.232702} \\
$^{6}$Li+$^{197}$Au & 27.79 & 1.54 & 7.51 & 0 & 0.557 & 1.470 & EvR & \cite{PhysRevC.89.024607} \\
$^{6}$Li+$^{209}$Bi & 29.87 & 1.84 & 8.71 & 0.104 & 0.168 & 0.451 & EvR+FF & \cite{PhysRevC.66.041602} \\
$^{6}$Li+$^{232}$Th & 31.61 & 2.76 & 11.28 & 0.156 & 0.206 & 0.532 & FF & \cite{PhysRevC.12.42} \\
$^{6}$Li+$^{238}$U & 30.93 & 2.62 & 10.33 & 0.081 & 0.685 & 1.760 & FF & \cite{PhysRevC.12.42} \\
$^{7}$Li+$^{16}$O & 4.31 & 0.89 & 8.99 & 0.029 & 0.677 & 0.669 & EvR & \cite{scholz1986complete} \\
$^{7}$Li+$^{28}$Si & 6.98 & 0.78 & 7.79 & 0 & 0.596 & 0.646 & EvR & \cite{pakou2009total} \\
$^{7}$Li+$^{59}$Co & 11.77 & 1.40 & 7.86 & 0 & 0.271 & 0.268 & EvR & \cite{PhysRevC.67.054602} \\
$^{7}$Li+$^{64}$Zn & 12.60 & 1.26 & 9.09 & 0.181 & 0.644 & 4.213 & EvR & \cite{PhysRevC.87.064614} \\
$^{7}$Li+$^{144}$Sm & 24.66 & 1.33 & 8.55 & 1.123 & 0.308 & 0.115 & EvR & \cite{PhysRevC.88.044617} \\
$^{7}$Li+$^{152}$Sm & 24.23 & 1.96 & 8.61 & 0 & 1.176 & 0.555 & EvR & \cite{PhysRevC.88.044617} \\
$^{7}$Li+$^{159}$Tb & 23.06 & 1.51 & 10.31 & 0.280 & 1.044 & 4.231 & EvR & \cite{BRODA1975356} \\
$^{7}$Li+$^{198}$Pt & 28.06 & 1.66 & 9.52 & 0 & 0.108 & 0.363 & EvR & \cite{SHRIVASTAVA2013931} \\
$^{7}$Li+$^{197}$Au & 28.79 & 1.80 & 10.32 & 0 & 0.019 & 0.044 & EvR & \cite{PhysRevC.89.024607} \\
$^{7}$Li+$^{209}$Bi & 29.69 & 1.69 & 9.86 & 0.149 & 0.302 & 1.279 & EvR+FF & \cite{PhysRevC.10.245} \\
$^{7}$Li+$^{209}$Bi & 29.53 & 1.67 & 9.58 & 0.188 & 0.056 & 0.072 & EvR+FF & \cite{PhysRevC.66.041602} \\
$^{7}$Li+$^{232}$Th & 31.12 & 2.12 & 11.00 & 0.120 & 0.366 & 6.620 & FF & \cite{PhysRevC.12.42} \\
$^{7}$Li+$^{235}$U & 31.59 & 2.13 & 12.23 & 0.337 & 0.354 & 0.136 & FF & \cite{PhysRevC.90.014603} \\
$^{7}$Li+$^{238}$U & 31.20 & 2.12 & 11.02 & 0.138 & 2.202 & 6.573 & FF & \cite{PhysRevC.12.42} \\
$^{7}$Be+$^{58}$Ni & 16.27 & 2.07 & 9.33 & 0 & 0.005 & 0.040 & EvR & \cite{PhysRevC.90.014616} \\
$^{7}$Be+$^{238}$U & 43.15 & 1.72 & 9.51 & 0.409 & 0.305 & 28.512 & FF & \cite{PhysRevC.74.044606} \\
$^{9}$Be+$^{89}$Y & 21.00 & 1.25 & 8.34 & 0.111 & 3.127 & 1.782 & EvR & \cite{PhysRevC.82.044608} \\
$^{9}$Be+$^{124}$Sn & 26.05 & 2.10 & 8.88 & 0 & 1.148 & 0.152 & EvR & \cite{PhysRevC.82.054601} \\
$^{9}$Be+$^{144}$Sm & 31.45 & 1.56 & 10.03 & 0 & 0.145 & 0.711 & EvR & \cite{PhysRevC.73.064606} \\
$^{9}$Be+$^{169}$Tm & 34.71 & 1.18 & 9.87 & 2.380 & 0.777 & 0.490 & EvR & \cite{PhysRevC.91.014608} \\
$^{9}$Be+$^{181}$Ta & 35.95 & 2.01 & 9.09 & 0.466 & 0.471 & 0.193 & EvR & \cite{PhysRevC.90.024621} \\
$^{9}$Be+$^{187}$Re & 37.01 & 2.08 & 9.92 & 0 & 0.346 & 0.439 & EvR & \cite{PhysRevC.91.014608} \\
$^{9}$Be+$^{208}$Pb & 38.32 & 1.77 & 9.31 & 0 & 1.146 & 0.803 & EvR+FF & \cite{PhysRevC.70.024606} \\
$^{9}$Be+$^{209}$Bi & 38.14 & 1.58 & 9.82 & 0.050 & 0.179 & 0.175 & EvR & \cite{liu2005partial} \\
$^{9}$Be+$^{209}$Bi & 38.15 & 1.56 & 10.01 & 0.107 & 0.199 & 0.466 & EvR & \cite{signorini1999does} \\
$^{9}$Be+$^{209}$Bi & 37.66 & 0.91 & 9.02 & 2.011 & 0.045 & 0.037 & EvR & \cite{signorini1998fusion} \\
$^{10}$B+$^{159}$Tb & 39.23 & 1.15 & 9.46 & 3.048 & 0.071 & 0.172 & EvR & \cite{MUKHERJEE200691} \\
$^{11}$B+$^{13}$C & 4.94 & 0.52 & 7.98 & 0.895 & 0.182 & 0.460 & EvR & \cite{DASMAHAPATRA1983192} \\
$^{11}$B+$^{159}$Tb & 39.41 & 2.20 & 10.31 & 0 & 0.039 & 0.171 & EvR & \cite{MUKHERJEE200691} \\
$^{11}$B+$^{238}$U & 49.66 & 1.64 & 11.34 & 0.333 & 0.181 & 0.719 & FF & \cite{liu1996fission} \\
$^{11}$B+$^{237}$Np & 54.84 & 2.79 & 11.80 & 0 & 0.312 & 1.132 & FF & \cite{liu1996fission} \\
$^{12}$C+$^{9}$Be & 3.86 & 0.63 & 7.12 & 0 & 0.237 & 1.023 & EvR & \cite{CHEUNG1978333} \\
$^{12}$C+$^{11}$B & 4.79 & 0.50 & 5.45 & 0.657 & 0.303 & 0.752 & EvR & \cite{HIGH1977149} \\
$^{12}$C+$^{11}$B & 4.88 & 0.50 & 6.30 & 0.880 & 1.386 & 0.920 & EvR & \cite{PhysRevLett.37.888} \\
$^{12}$C+$^{13}$C & 5.63 & 0.52 & 6.82 & 0.558 & 0.495 & 0.688 & EvR & \cite{PhysRevLett.37.888} \\
$^{12}$C+$^{14}$C & 5.49 & 0.39 & 6.24 & 0 & 0.002 & 0.006 & EvR & \cite{DASMAHAPATRA1993657} \\
$^{12}$C+$^{14}$N & 6.90 & 0.64 & 7.64 & 0.646 & 0.424 & 0.424 & EvR & \cite{PhysRevLett.37.888} \\
$^{12}$C+$^{20}$Ne & 9.80 & 0.88 & 7.66 & 0.170 & 1.405 & 1.692 & EvR & \cite{hulke1980comparison} \\
$^{12}$C+$^{46}$Ti & 21.31 & 1.58 & 9.82 & 0.118 & 0.222 & 0.911 & EvR & \cite{BOZEK1986171} \\
$^{12}$C+$^{48}$Ti & 20.40 & 1.53 & 8.29 & 0 & 0.486 & 0.728 & EvR & \cite{BOZEK1986171} \\
$^{12}$C+$^{50}$Ti & 19.72 & 1.20 & 8.12 & 0.296 & 0.167 & 0.409 & EvR & \cite{BOZEK1986171} \\
$^{12}$C+$^{89}$Y & 32.17 & 1.50 & 10.50 & 0.202 & 0.256 & 0.830 & EvR & \cite{PhysRevC.82.044608} \\
$^{12}$C+$^{92}$Zr & 31.82 & 1.25 & 9.13 & 0 & 3.264 & 13.703 & EvR & \cite{PhysRevC.64.064608} \\
$^{12}$C+$^{144}$Sm & 46.83 & 1.54 & 11.63 & 0.127 & 0.501 & 1.100 & EvR & \cite{PhysRevC.46.244} \\
$^{12}$C+$^{152}$Sm & 47.22 & 2.09 & 11.20 & 2.101 & 0.043 & 0.024 & EvR & \cite{BRODA1975356} \\
$^{12}$C+$^{154}$Sm & 44.54 & 0.84 & 9.40 & 1.831 & 0.144 & 0.302 & EvR & \cite{PhysRevC.31.1752} \\
$^{12}$C+$^{181}$Ta & 52.43 & 2.07 & 10.66 & 0 & 0.211 & 0.216 & EvR & \cite{crippa1994excitation} \\
$^{12}$C+$^{194}$Pt & 54.91 & 1.52 & 10.34 & 0.127 & 1.115 & 0.363 & EvR+FF & \cite{PhysRevC.63.054602} \\
$^{12}$C+$^{198}$Pt & 55.27 & 1.55 & 10.62 & 0.161 & 2.152 & 0.222 & EvR+FF & \cite{PhysRevC.63.054602} \\
$^{12}$C+$^{204}$Pb & 55.99 & 1.04 & 11.23 & 0.190 & 0.482 & 2.494 & EvR+FF & \cite{PhysRevC.68.014603} \\
$^{12}$C+$^{208}$Pb & 56.34 & 1.26 & 10.59 & 0 & 1.329 & 0.091 & EvR+FF & \cite{PhysRevC.75.044608} \\
$^{12}$C+$^{237}$Np & 62.96 & 2.52 & 10.51 & 0 & 0.326 & 1.852 & FF & \cite{liu1996fission} \\
$^{13}$C+$^{10}$B & 4.94 & 0.63 & 7.67 & 0 & 0.124 & 0.190 & EvR & \cite{DASMAHAPATRA1983192} \\
$^{13}$C+$^{11}$B & 5.02 & 0.54 & 8.44 & 0.864 & 0.200 & 0.212 & EvR & \cite{DASMAHAPATRA1983192} \\
$^{13}$C+$^{13}$C & 5.98 & 0.64 & 8.08 & 0.335 & 34.259 & 0.956 & EvR & \cite{CHATTERJEE1980273} \\
$^{13}$C+$^{48}$Ti & 20.50 & 2.45 & 8.64 & 0 & 0.074 & 0.190 & EvR & \cite{DUMONT1985301} \\
$^{13}$C+$^{232}$Th & 61.68 & 2.20 & 12.75 & 2.002 & 0.198 & 0.036 & FF & \cite{PhysRevC.72.067601} \\
$^{14}$N+$^{10}$B & 5.72 & 0.65 & 7.79 & 0.153 & 0.180 & 1.105 & EvR & \cite{SHIUCHINWU1978177} \\
$^{14}$N+$^{12}$C & 6.90 & 0.64 & 7.64 & 0.648 & 0.430 & 0.430 & EvR & \cite{SWITKOWSKI1977502} \\
$^{14}$N+$^{14}$N & 7.55 & 0.74 & 6.10 & 0 & 0.024 & 0.042 & EvR & \cite{PhysRevC.26.1482} \\
$^{14}$N+$^{14}$N & 8.25 & 0.81 & 8.12 & 0.181 & 2.372 & 1.327 & EvR & \cite{SWITKOWSKI1976202} \\
$^{14}$N+$^{14}$N & 8.23 & 0.80 & 8.11 & 0.196 & 1.537 & 1.370 & EvR & \cite{PhysRevLett.37.888} \\
$^{14}$N+$^{16}$O & 9.23 & 0.83 & 8.70 & 0.199 & 0.221 & 0.228 & EvR & \cite{PhysRevLett.37.888} \\
$^{14}$N+$^{59}$Co & 26.81 & 1.41 & 9.46 & 0.367 & 0.282 & 0.509 & EvR & \cite{GOMES1989395} \\
$^{14}$N+$^{232}$Th & 70.99 & 2.03 & 12.13 & 1.933 & 0.185 & 0.163 & FF & \cite{PhysRevC.69.064603} \\
$^{15}$N+$^{54}$Fe & 26.19 & 1.81 & 10.04 & 0 & 0.554 & 0.952 & EvR & \cite{FUNAKI1993307} \\
$^{15}$N+$^{209}$Bi & 66.53 & 1.58 & 10.90 & 0 & 0.159 & 1.595 & FF & \cite{PhysRevC.33.2017} \\
$^{16}$O+$^{12}$C & 7.72 & 0.67 & 7.81 & 0.543 & 4.989 & 1.782 & EvR & \cite{CHRISTENSEN1977189} \\
$^{16}$O+$^{13}$C & 7.80 & 0.80 & 7.29 & 0 & 0.344 & 0.590 & EvR & \cite{DASMAHAPATRA1991395} \\
$^{16}$O+$^{14}$N & 9.16 & 0.82 & 8.25 & 0.176 & 0.259 & 0.268 & EvR & \cite{SWITKOWSKI1977502} \\
$^{16}$O+$^{16}$O & 10.34 & 0.92 & 8.50 & 0.146 & 0.253 & 0.250 & EvR & \cite{PhysRevC.35.591} \\
$^{16}$O+$^{16}$O & 10.32 & 0.95 & 9.09 & 0 & 0.229 & 0.638 & EvR & \cite{PhysRevC.33.1679} \\
$^{16}$O+$^{27}$Al & 15.76 & 1.19 & 7.41 & 0 & 0.015 & 0.077 & EvR & \cite{DAUK1975170} \\
$^{16}$O+$^{46}$Ti & 26.21 & 1.40 & 8.92 & 0.459 & 0.071 & 0.076 & EvR & \cite{NETO1990333} \\
$^{16}$O+$^{50}$Ti & 26.06 & 1.65 & 8.28 & 0.050 & 0.222 & 1.031 & EvR & \cite{GOMES1989395} \\
$^{16}$O+$^{50}$Ti & 25.98 & 1.55 & 9.12 & 0.125 & 0.075 & 0.445 & EvR & \cite{NETO1990333} \\
$^{16}$O+$^{54}$Fe & 30.29 & 1.46 & 8.76 & 0.404 & 0.340 & 0.693 & EvR & \cite{FUNAKI1993307} \\
$^{16}$O+$^{56}$Fe & 30.24 & 1.05 & 9.13 & 3.085 & 0.445 & 0.330 & EvR & \cite{FUNAKI1993307} \\
$^{16}$O+$^{58}$Ni & 31.32 & 1.07 & 9.33 & 0.128 & 1.260 & 1.556 & EvR & \cite{KEELEY19981} \\
$^{16}$O+$^{62}$Ni & 30.54 & 1.04 & 8.93 & 0.179 & 1.265 & 0.472 & EvR & \cite{KEELEY19981} \\
$^{16}$O+$^{63}$Cu & 34.88 & 2.77 & 9.20 & 0 & 0.391 & 0.390 & EvR & \cite{CHAMON199229} \\
$^{16}$O+$^{63}$Cu & 35.32 & 2.89 & 9.44 & 0 & 0.634 & 0.518 & EvR & \cite{PEREIRA1989347} \\
$^{16}$O+$^{63}$Cu & 33.07 & 1.03 & 8.95 & 1.402 & 0.115 & 0.003 & EvR & \cite{PhysRevC.14.152} \\
$^{16}$O+$^{65}$Cu & 33.68 & 2.86 & 8.97 & 0 & 0.302 & 0.358 & EvR & \cite{CHAMON199229} \\
$^{16}$O+$^{64}$Zn & 33.00 & 2.54 & 10.32 & 0 & 0.230 & 0.265 & EvR & \cite{PhysRevC.71.034608} \\
$^{16}$O+$^{70}$Ge & 34.76 & 1.45 & 9.75 & 0 & 0.955 & 0.189 & EvR & \cite{PhysRevC.52.3103} \\
$^{16}$O+$^{72}$Ge & 35.62 & 1.79 & 10.03 & 0 & 1.700 & 0.370 & EvR & \cite{PhysRevC.52.3103} \\
$^{16}$O+$^{73}$Ge & 33.98 & 1.01 & 8.31 & 1.880 & 0.064 & 0.034 & EvR & \cite{PhysRevC.52.3103} \\
$^{16}$O+$^{74}$Ge & 34.89 & 1.16 & 9.66 & 2.095 & 1.729 & 0.631 & EvR & \cite{PhysRevC.52.3103} \\
$^{16}$O+$^{76}$Ge & 34.95 & 1.12 & 9.43 & 1.966 & 9.775 & 0.612 & EvR & \cite{PhysRevC.86.044621} \\
$^{16}$O+$^{76}$Ge & 34.50 & 1.13 & 9.55 & 1.624 & 1.188 & 0.279 & EvR & \cite{PhysRevC.52.3103} \\
$^{16}$O+$^{92}$Zr & 41.47 & 1.58 & 9.56 & 0 & 3.633 & 1.394 & EvR & \cite{PhysRevC.64.064608} \\
$^{16}$O+$^{112}$Cd & 48.28 & 1.66 & 11.02 & 0 & 0.414 & 1.891 & EvR & \cite{ACKERMANN1994374} \\
$^{16}$O+$^{112}$Sn & 50.61 & 1.38 & 9.81 & 0.127 & 0.811 & 19.177 & EvR & \cite{PhysRevC.65.014614} \\
$^{16}$O+$^{116}$Sn & 50.20 & 1.39 & 9.96 & 0.155 & 0.793 & 3.315 & EvR & \cite{PhysRevC.65.014614} \\
$^{16}$O+$^{144}$Nd & 57.31 & 1.53 & 11.12 & 0 & 0.042 & 0.145 & EvR & \cite{PhysRevC.47.2970} \\
$^{16}$O+$^{150}$Nd & 57.04 & 1.67 & 9.25 & 0 & 0.555 & 0.120 & EvR & \cite{BRODA1975356} \\
$^{16}$O+$^{144}$Sm & 60.51 & 1.59 & 10.34 & 0 & 21.851 & 4.572 & EvR & \cite{PhysRevC.52.3151} \\
$^{16}$O+$^{147}$Sm & 58.84 & 1.44 & 9.61 & 0 & 0.797 & 1.554 & EvR & \cite{PhysRevC.39.516} \\
$^{16}$O+$^{148}$Sm & 59.73 & 1.75 & 10.51 & 1.158 & 9.033 & 1.269 & EvR & \cite{PhysRevC.52.3151} \\
$^{16}$O+$^{148}$Sm & 59.74 & 2.07 & 11.26 & 0 & 0.215 & 0.384 & EvR & \cite{PhysRevC.39.516} \\
$^{16}$O+$^{148}$Sm & 59.15 & 1.88 & 9.73 & 0 & 0.154 & 0.154 & EvR & \cite{PhysRevC.21.2427} \\
$^{16}$O+$^{149}$Sm & 58.95 & 1.88 & 9.97 & 0 & 0.523 & 1.714 & EvR & \cite{PhysRevC.39.516} \\
$^{16}$O+$^{150}$Sm & 59.20 & 1.67 & 10.64 & 1.481 & 0.099 & 0.098 & EvR & \cite{PhysRevC.21.2427} \\
$^{16}$O+$^{152}$Sm & 59.67 & 1.86 & 10.86 & 2.197 & 0.216 & 0.204 & EvR & \cite{PhysRevC.21.2427} \\
$^{16}$O+$^{154}$Sm & 58.90 & 2.53 & 10.13 & 0 & 0.039 & 0.037 & EvR & \cite{PhysRevC.21.2427} \\
$^{16}$O+$^{154}$Sm & 59.03 & 2.80 & 10.30 & 0 & 0.095 & 0.024 & EvR & \cite{jahnke1982global} \\
$^{16}$O+$^{154}$Sm & 59.20 & 1.99 & 10.46 & 1.543 & 3.208 & 1.783 & EvR & \cite{PhysRevC.52.3151} \\
$^{16}$O+$^{154}$Sm & 59.21 & 1.84 & 10.20 & 1.992 & 3.395 & 2.101 & EvR & \cite{PhysRevLett.67.3368} \\
$^{16}$O+$^{166}$Er & 63.66 & 1.48 & 10.29 & 2.643 & 0.673 & 1.126 & EvR & \cite{PhysRevC.43.2303} \\
$^{16}$O+$^{174}$Yb & 66.52 & 2.59 & 10.93 & 0.119 & 0.639 & 1.329 & EvR & \cite{PhysRevC.93.054622} \\
$^{16}$O+$^{176}$Yb & 66.79 & 2.94 & 9.75 & 0.100 & 1.420 & 4.830 & EvR & \cite{PhysRevC.93.054622} \\
$^{16}$O+$^{186}$W  & 68.31 & 2.29 & 10.44 & 0 & 6.333 & 1.323 & EvR+FF & \cite{LEMMON199332} \\
$^{16}$O+$^{186}$W  & 68.18 & 2.13 & 10.38 & 0.106 & 0.383 & 1.874 & EvR+FF & \cite{trotta2005fusion} \\
$^{16}$O+$^{194}$Pt & 72.60 & 2.18 & 11.73 & 0.003 & 0.028 & 0.047 & EvR+FF & \cite{PRASAD201262,PhysRevC.84.064606} \\
$^{16}$O+$^{208}$Pb & 73.79 & 1.20 & 10.62 & 1.877 & 29.330 & 0.089 & EvR+FF & \cite{PhysRevC.60.044608} \\
$^{16}$O+$^{208}$Pb & 74.13 & 1.23 & 10.59 & 2.135 & 0.097 & 0.151 & EvR+FF & \cite{PhysRevC.33.2017} \\
$^{16}$O+$^{209}$Bi & 75.03 & 1.21 & 10.75 & 2.245 & 0.088 & 0.138 & FF & \cite{PhysRevC.33.2017} \\
$^{16}$O+$^{209}$Bi & 77.17 & 2.43 & 10.11 & 0.219 & 1.923 & 2.397 & FF & \cite{PhysRev.135.B669} \\
$^{17}$O+$^{12}$C & 7.66 & 0.70 & 7.27 & 0.208 & 0.111 & 0.065 & EvR & \cite{PhysRevC.13.1527} \\
$^{17}$O+$^{12}$C & 9.50 & 0.83 & 11.51 & 3.671 & 1.253 & 2.932 & EvR & \cite{PhysRevC.47.2699} \\
$^{17}$O+$^{13}$C & 7.88 & 0.63 & 6.81 & 0.080 & 0.275 & 0.432 & EvR & \cite{PhysRevC.47.2699} \\
$^{17}$O+$^{16}$O & 10.08 & 0.93 & 9.11 & 0.143 & 0.628 & 11.320 & EvR & \cite{PhysRevC.33.1679} \\
$^{17}$O+$^{144}$Sm & 60.25 & 1.86 & 10.42 & 0 & 5.928 & 4.610 & EvR & \cite{PhysRevC.52.3151} \\
$^{18}$O+$^{9}$Be & 4.96 & 0.49 & 6.57 & 0.858 & 0.090 & 1.765 & EvR & \cite{ROTH1980148} \\
$^{18}$O+$^{12}$C & 7.43 & 0.35 & 7.34 & 2.401 & 0.052 & 0.028 & EvR & \cite{PhysRevC.13.1527} \\
$^{18}$O+$^{12}$C & 7.71 & 0.87 & 7.42 & 0 & 0.608 & 0.729 & EvR & \cite{PhysRevC.90.041603} \\
$^{18}$O+$^{16}$O & 9.90 & 0.85 & 8.07 & 0.153 & 0.419 & 0.769 & EvR & \cite{PhysRevC.33.1679} \\
$^{18}$O+$^{44}$Ca & 22.44 & 1.37 & 8.35 & 0.027 & 0.230 & 0.643 & EvR & \cite{BOZEK1986171} \\
$^{18}$O+$^{58}$Ni & 31.68 & 2.85 & 8.02 & 0 & 2.540 & 0.995 & EvR & \cite{PhysRevC.46.2360} \\
$^{18}$O+$^{60}$Ni & 33.77 & 2.72 & 10.70 & 0 & 1.474 & 1.493 & EvR & \cite{PhysRevC.55.3155} \\
$^{18}$O+$^{64}$Ni & 33.29 & 2.36 & 9.77 & 0 & 0.456 & 0.488 & EvR & \cite{PhysRevC.55.3155} \\
$^{18}$O+$^{63}$Cu & 34.32 & 2.30 & 10.09 & 0.513 & 0.229 & 0.186 & EvR & \cite{CHAMON199229} \\
$^{18}$O+$^{74}$Ge & 34.31 & 1.09 & 9.23 & 2.076 & 32.807 & 0.795 & EvR & \cite{PhysRevC.86.044621} \\
$^{18}$O+$^{148}$Nd & 58.21 & 2.95 & 9.13 & 0 & 0.088 & 0.101 & EvR & \cite{BRODA1975356} \\
$^{18}$O+$^{208}$Pb & 74.14 & 1.33 & 10.79 & 2.562 & 0.175 & 0.290 & FF & \cite{PhysRevC.33.2017} \\
$^{19}$F+$^{54}$Fe & 33.87 & 3.14 & 9.50 & 0 & 0.334 & 0.564 & EvR & \cite{FUNAKI1993307} \\
$^{19}$F+$^{56}$Fe & 33.00 & 2.73 & 9.10 & 0 & 0.611 & 0.852 & EvR & \cite{FUNAKI1993307} \\
$^{19}$F+$^{93}$Nb & 47.66 & 2.31 & 9.98 & 0 & 0.965 & 0.217 & EvR & \cite{prasad1996study} \\
$^{19}$F+$^{181}$Ta & 76.22 & 2.90 & 10.96 & 0.954 & 0.149 & 0.031 & EvR+FF & \cite{HINDE1982109} \\
$^{19}$F+$^{188}$Os & 78.21 & 2.96 & 10.34 & 0 & 0.208 & 0.259 & EvR+FF & \cite{MAHATA2003209} \\
$^{19}$F+$^{192}$Os & 78.80 & 2.11 & 10.31 & 3.142 & 0.028 & 0.028 & EvR+FF & \cite{MAHATA2003209} \\
$^{19}$F+$^{208}$Pb & 82.48 & 2.28 & 10.97 & 0.516 & 12.345 & 0.161 & EvR+FF & \cite{PhysRevC.60.054602} \\
$^{19}$F+$^{208}$Pb & 83.47 & 2.71 & 11.22 & 0.275 & 0.211 & 0.294 & FF & \cite{HUANQIAO1990531} \\
$^{19}$F+$^{208}$Pb & 83.04 & 2.57 & 12.87 & 0.066 & 0.772 & 3.277 & FF & \cite{PhysRevLett.81.3341} \\
$^{19}$F+$^{209}$Bi & 83.96 & 2.60 & 10.84 & 0 & 0.722 & 0.337 & FF & \cite{samant2000fission} \\
$^{19}$F+$^{232}$Th & 87.85 & 3.90 & 9.91 & 0 & 0.698 & 0.732 & FF & \cite{PhysRevC.51.3109} \\
$^{19}$F+$^{232}$Th & 89.94 & 4.23 & 12.01 & 0 & 0.034 & 0.071 & FF & \cite{PhysRevC.43.1466} \\
$^{19}$F+$^{232}$Th & 90.28 & 4.61 & 13.51 & 0.373 & 1.815 & 6.393 & FF & \cite{ZHANG1989133} \\
$^{20}$Ne+$^{208}$Pb & 94.52 & 2.16 & 11.22 & 2.121 & 0.600 & 1.025 & FF & \cite{PhysRevC.85.054608} \\
$^{20}$Ne+$^{238}$U & 102.82 & 5.08 & 12.52 & 0 & 0.703 & 0.933 & FF & \cite{PhysRev.128.767} \\
$^{23}$Na+$^{48}$Ti & 33.45 & 0.86 & 9.30 & 3.361 & 0.199 & 0.365 & EvR & \cite{PhysRevLett.57.2002} \\
$^{23}$Na+$^{206}$Pb & 99.46 & 2.74 & 11.59 & 0.002 & 0.002 & 0.006 & FF & \cite{PhysRevLett.57.2002} \\
$^{24}$Mg+$^{24}$Mg & 22.07 & 1.09 & 8.57 & 1.437 & 1.341 & 0.194 & EvR & \cite{PhysRevC.25.1877} \\
$^{24}$Mg+$^{26}$Mg & 20.89 & 1.24 & 8.35 & 0 & 0.376 & 0.047 & EvR & \cite{PhysRevC.25.1877} \\
$^{24}$Mg+$^{30}$Si & 24.10 & 1.06 & 8.07 & 0.107 & 0.260 & 1.108 & EvR & \cite{PhysRevLett.113.022701} \\
$^{26}$Mg+$^{248}$Cm & 128.60 & 5.16 & 14.09 & 0.313 & 0.043 & 1.212 & FF & \cite{itkis2010fusion} \\
$^{27}$Al+$^{45}$Sc & 37.86 & 1.33 & 7.70 & 0.237 & 0.102 & 1.927 & EvR & \cite{PhysRevC.81.024611} \\
$^{27}$Al+$^{70}$Ge & 54.15 & 1.76 & 9.08 & 0 & 1.856 & 0.719 & EvR & \cite{PhysRevC.41.910} \\
$^{27}$Al+$^{72}$Ge & 54.00 & 1.61 & 9.02 & 0.238 & 1.700 & 0.750 & EvR & \cite{PhysRevC.41.910} \\
$^{27}$Al+$^{73}$Ge & 54.12 & 1.60 & 8.79 & 1.254 & 0.485 & 0.363 & EvR & \cite{PhysRevC.41.910} \\
$^{27}$Al+$^{74}$Ge & 53.15 & 1.20 & 8.14 & 1.475 & 0.846 & 0.689 & EvR & \cite{PhysRevC.41.910} \\
$^{27}$Al+$^{76}$Ge & 53.17 & 1.33 & 8.87 & 1.220 & 1.028 & 0.428 & EvR & \cite{PhysRevC.41.910} \\
$^{29}$Al+$^{197}$Au & 111.09 & 2.96 & 11.00 & 0 & 0.090 & 0.528 & FF & \cite{watanabe2001measurement} \\
$^{28}$Si+$^{24}$Mg & 24.51 & 1.09 & 8.13 & 0.082 & 1.035 & 15.515 & EvR & \cite{PhysRevC.41.988} \\
$^{28}$Si+$^{24}$Mg & 24.64 & 0.91 & 8.10 & 1.318 & 1.536 & 0.160 & EvR & \cite{PhysRevC.25.1877} \\
$^{28}$Si+$^{26}$Mg & 24.91 & 1.10 & 8.47 & 0.110 & 1.834 & 1.834 & EvR & \cite{PhysRevC.41.988} \\
$^{28}$Si+$^{28}$Si & 29.03 & 1.63 & 8.26 & 0 & 2.384 & 0.919 & EvR & \cite{PhysRevC.25.1877} \\
$^{28}$Si+$^{28}$Si & 29.53 & 1.41 & 9.12 & 0.207 & 0.805 & 25.466 & EvR & \cite{PhysRevC.90.044608} \\
$^{28}$Si+$^{29}$Si & 28.53 & 1.42 & 8.24 & 0 & 2.245 & 1.188 & EvR & \cite{PhysRevC.25.1877} \\
$^{28}$Si+$^{30}$Si & 28.22 & 1.41 & 8.41 & 0 & 1.645 & 0.270 & EvR & \cite{PhysRevC.25.1877} \\
$^{28}$Si+$^{30}$Si & 28.73 & 1.75 & 8.18 & 0 & 0.158 & 0.441 & EvR & \cite{BOZEK1986171} \\
$^{28}$Si+$^{30}$Si & 28.19 & 1.13 & 8.02 & 0.120 & 1.054 & 1.217 & EvR & \cite{PhysRevC.78.017601} \\
$^{28}$Si+$^{58}$Ni & 53.92 & 1.52 & 9.01 & 0.188 & 3.051 & 8.115 & EvR & \cite{PhysRevC.30.2088} \\
$^{28}$Si+$^{62}$Ni & 51.33 & 1.22 & 7.86 & 0 & 0.254 & 2.776 & EvR & \cite{STEFANINI1986509} \\
$^{28}$Si+$^{64}$Ni & 51.33 & 0.99 & 8.51 & 2.686 & 0.167 & 0.910 & EvR & \cite{STEFANINI1986509} \\
$^{28}$Si+$^{64}$Ni & 50.37 & 1.30 & 7.23 & 0.131 & 0.328 & 1.643 & EvR & \cite{JIANG200618} \\
$^{28}$Si+$^{64}$Ni & 50.72 & 1.23 & 7.18 & 0 & 0.100 & 0.584 & EvR & \cite{A_K_Sinha_1997} \\
$^{28}$Si+$^{68}$Zn & 53.43 & 1.40 & 7.47 & 0 & 4.352 & 3.577 & EvR & \cite{DASGUPTA1992351} \\
$^{28}$Si+$^{68}$Zn & 53.42 & 1.44 & 7.67 & 0 & 0.350 & 2.250 & EvR & \cite{PhysRevLett.66.1414} \\
$^{28}$Si+$^{90}$Zr & 72.36 & 2.17 & 10.24 & 0.110 & 0.795 & 1.751 & EvR & \cite{PhysRevC.81.044610} \\
$^{28}$Si+$^{92}$Zr & 70.23 & 2.47 & 9.63 & 0 & 5.736 & 0.540 & EvR & \cite{PhysRevC.64.064608} \\
$^{28}$Si+$^{94}$Zr & 69.70 & 1.94 & 8.25 & 0.208 & 1.532 & 5.135 & EvR & \cite{PhysRevC.81.044610} \\
$^{28}$Si+$^{96}$Zr & 70.00 & 1.81 & 9.75 & 2.990 & 0.501 & 1.160 & EvR & \cite{PhysRevC.96.014614} \\
$^{28}$Si+$^{93}$Nb & 73.52 & 1.63 & 10.45 & 2.972 & 0.176 & 0.733 & EvR & \cite{PhysRevC.56.1936} \\
$^{28}$Si+$^{94}$Mo & 74.20 & 1.26 & 8.03 & 2.142 & 2.164 & 213.559 & EvR & \cite{ACKERMANN199691} \\
$^{28}$Si+$^{94}$Mo & 75.35 & 1.75 & 8.89 & 1.487 & 0.239 & 1.190 & EvR & \cite{ACKERMANN1995129} \\
$^{28}$Si+$^{144}$Nd & 101.95 & 2.64 & 11.94 & 2.348 & 0.421 & 0.244 & EvR & \cite{A_K_Sinha_1997} \\
$^{28}$Si+$^{154}$Sm & 101.11 & 4.85 & 10.58 & 0 & 1.434 & 2.274 & EvR & \cite{PhysRevLett.65.3100} \\
$^{28}$Si+$^{164}$Er & 107.01 & 3.50 & 11.66 & 2.516 & 0.038 & 0.051 & EvR+FF & \cite{HINDE1986550} \\
$^{28}$Si+$^{170}$Er & 104.15 & 2.69 & 11.31 & 3.969 & 0.032 & 0.061 & EvR+FF & \cite{HINDE1986550} \\
$^{28}$Si+$^{178}$Hf & 114.59 & 2.98 & 10.75 & 2.453 & 7.061 & 0.197 & EvR+FF & \cite{PhysRevC.66.044601} \\
$^{28}$Si+$^{198}$Pt & 121.64 & 2.54 & 10.28 & 2.171 & 0.276 & 0.618 & FF & \cite{PhysRevC.62.014602} \\
$^{28}$Si+$^{208}$Pb & 126.74 & 2.15 & 10.88 & 3.212 & 0.147 & 0.150 & FF & \cite{HINDE1995271} \\
$^{29}$Si+$^{178}$Hf & 114.73 & 4.36 & 11.15 & 0 & 11.560 & 2.254 & EvR+FF & \cite{PhysRevC.66.044601} \\
$^{30}$Si+$^{24}$Mg & 23.98 & 1.01 & 8.07 & 0 & 0.326 & 0.293 & EvR & \cite{PhysRevC.41.988} \\
$^{30}$Si+$^{26}$Mg & 24.87 & 1.14 & 9.51 & 0.140 & 2.310 & 4.394 & EvR & \cite{PhysRevC.41.988} \\
$^{30}$Si+$^{30}$Si & 28.10 & 0.91 & 8.60 & 0 & 0.763 & 1.862 & EvR & \cite{BOZEK1986171} \\
$^{30}$Si+$^{58}$Ni & 52.74 & 1.44 & 8.70 & 0.277 & 0.220 & 1.676 & EvR & \cite{STEFANINI1986509} \\
$^{30}$Si+$^{62}$Ni & 51.94 & 1.42 & 9.57 & 0.180 & 0.552 & 7.554 & EvR & \cite{STEFANINI1986509} \\
$^{30}$Si+$^{64}$Ni & 51.29 & 1.31 & 9.45 & 0.190 & 0.404 & 0.902 & EvR & \cite{STEFANINI1986509} \\
$^{30}$Si+$^{170}$Er & 109.34 & 3.40 & 11.61 & 2.020 & 0.087 & 0.180 & EvR+FF & \cite{HINDE1982109} \\
$^{30}$Si+$^{238}$U & 137.63 & 5.38 & 13.12 & 0.271 & 0.228 & 5.523 & FF & \cite{nishio2006measurement} \\
$^{30}$Si+$^{238}$U & 137.69 & 4.68 & 11.07 & 0 & 0.285 & 6.054 & FF & \cite{PhysRevC.82.044604} \\
$^{31}$P+$^{175}$Lu & 120.53 & 4.75 & 10.98 & 0 & 2.505 & 0.171 & EvR+FF & \cite{PhysRevC.66.044601} \\
$^{32}$S+$^{12}$C & 15.14 & 1.07 & 8.30 & 0 & 0.291 & 1.010 & EvR & \cite{PhysRevC.41.2654} \\
$^{32}$S+$^{13}$C & 15.78 & 1.29 & 9.55 & 0.077 & 0.276 & 0.743 & EvR & \cite{PhysRevC.41.2654} \\
$^{32}$S+$^{24}$Mg & 27.84 & 1.12 & 8.71 & 0.489 & 0.118 & 0.178 & EvR & \cite{PhysRevC.28.667} \\
$^{32}$S+$^{25}$Mg & 27.09 & 1.11 & 8.31 & 0 & 0.256 & 0.378 & EvR & \cite{PhysRevC.28.667} \\
$^{32}$S+$^{26}$Mg & 27.05 & 1.19 & 8.47 & 0 & 0.305 & 0.820 & EvR & \cite{PhysRevC.28.667} \\
$^{32}$S+$^{27}$Al & 29.70 & 1.17 & 8.78 & 0 & 0.612 & 1.004 & EvR & \cite{PhysRevC.28.667} \\
$^{32}$S+$^{48}$Ca & 42.57 & 1.49 & 8.16 & 0.182 & 0.114 & 6.221 & EvR & \cite{PhysRevC.87.014611} \\
$^{32}$S+$^{58}$Ni & 58.47 & 0.89 & 8.21 & 1.724 & 1.002 & 2.401 & EvR & \cite{PhysRevC.42.1530} \\
$^{32}$S+$^{58}$Ni & 59.64 & 1.32 & 8.49 & 0.114 & 0.264 & 1.688 & EvR & \cite{STEFANINI1986509} \\
$^{32}$S+$^{58}$Ni & 59.60 & 1.29 & 8.38 & 0.156 & 0.672 & 2.871 & EvR & \cite{STEFANINI198566} \\
$^{32}$S+$^{64}$Ni & 56.74 & 1.03 & 8.38 & 2.005 & 0.314 & 0.154 & EvR & \cite{PhysRevC.42.1530} \\
$^{32}$S+$^{64}$Ni & 58.94 & 1.73 & 10.21 & 2.964 & 1.238 & 0.704 & EvR & \cite{DASGUPTA1992351} \\
$^{32}$S+$^{64}$Ni & 59.40 & 2.99 & 11.03 & 0 & 0.239 & 1.661 & EvR & \cite{PhysRevLett.66.1414} \\
$^{32}$S+$^{64}$Ni & 57.44 & 1.51 & 8.32 & 0.154 & 0.361 & 2.129 & EvR & \cite{STEFANINI198566} \\
$^{32}$S+$^{64}$Ni & 57.41 & 1.51 & 8.33 & 0.156 & 0.321 & 2.559 & EvR & \cite{STEFANINI1986509} \\
$^{32}$S+$^{89}$Y & 76.94 & 1.22 & 9.63 & 2.132 & 9.856 & 2.061 & EvR & \cite{PhysRevC.66.034607} \\
$^{32}$S+$^{90}$Zr & 79.37 & 1.68 & 10.40 & 2.180 & 44.555 & 4.001 & EvR & \cite{PhysRevC.82.054609} \\
$^{32}$S+$^{94}$Zr & 78.97 & 2.53 & 10.80 & 1.692 & 41.002 & 2.852 & EvR & \cite{PhysRevC.89.064605} \\
$^{32}$S+$^{96}$Zr & 78.20 & 2.31 & 10.11 & 1.949 & 42.939 & 1.809 & EvR & \cite{PhysRevC.82.054609} \\
$^{32}$S+$^{94}$Mo & 83.16 & 2.35 & 10.73 & 0 & 1.253 & 56.894 & EvR & \cite{PENGO1983255} \\
$^{32}$S+$^{98}$Mo & 82.16 & 1.89 & 9.20 & 2.777 & 5.571 & 13.037 & EvR & \cite{PENGO1983255} \\
$^{32}$S+$^{100}$Mo & 83.29 & 2.57 & 10.18 & 1.925 & 2.959 & 2.484 & EvR & \cite{PENGO1983255} \\
$^{32}$S+$^{100}$Ru & 84.05 & 1.88 & 8.32 & 0.182 & 0.218 & 8.962 & EvR & \cite{PENGO1983255} \\
$^{32}$S+$^{101}$Ru & 84.77 & 2.61 & 8.51 & 0 & 0.279 & 0.902 & EvR & \cite{PENGO1983255} \\
$^{32}$S+$^{102}$Ru & 84.04 & 2.31 & 8.59 & 0 & 0.615 & 2.061 & EvR & \cite{PENGO1983255} \\
$^{32}$S+$^{104}$Ru & 83.31 & 2.58 & 8.38 & 0 & 0.075 & 0.449 & EvR & \cite{PENGO1983255} \\
$^{32}$S+$^{103}$Rh & 85.24 & 1.71 & 7.33 & 1.083 & 1.754 & 6.753 & EvR & \cite{PENGO1983255} \\
$^{32}$S+$^{105}$Pd & 86.80 & 2.13 & 7.35 & 0 & 0.046 & 0.230 & EvR & \cite{PENGO1983255} \\
$^{32}$S+$^{106}$Pd & 86.00 & 1.80 & 7.28 & 0 & 0.304 & 2.938 & EvR & \cite{PENGO1983255} \\
$^{32}$S+$^{108}$Pd & 85.77 & 2.05 & 7.69 & 0 & 0.199 & 3.589 & EvR & \cite{PENGO1983255} \\
$^{32}$S+$^{110}$Pd & 87.01 & 2.15 & 8.61 & 1.395 & 11.141 & 0.390 & EvR & \cite{PhysRevC.52.R1727} \\
$^{32}$S+$^{110}$Pd & 87.82 & 2.18 & 8.59 & 1.863 & 0.275 & 0.583 & EvR & \cite{PENGO1983255} \\
$^{32}$S+$^{112}$Sn & 94.98 & 1.73 & 8.93 & 0.965 & 0.117 & 0.598 & EvR & \cite{PhysRevC.65.014614} \\
$^{32}$S+$^{120}$Sn & 94.13 & 1.89 & 9.65 & 1.738 & 0.255 & 0.331 & EvR & \cite{PhysRevC.65.014614} \\
$^{32}$S+$^{138}$Ba & 107.55 & 3.14 & 10.36 & 1.241 & 4.382 & 3.078 & EvR+FF & \cite{PhysRevC.51.1336} \\
$^{32}$S+$^{154}$Sm & 112.78 & 3.64 & 8.91 & 1.133 & 0.220 & 0.403 & EvR+FF & \cite{PhysRevC.49.245} \\
$^{32}$S+$^{182}$W  & 131.10 & 2.39 & 9.89 & 2.575 & 0.526 & 2.395 & FF & \cite{PhysRevC.62.054603} \\
$^{32}$S+$^{184}$W  & 127.12 & 2.91 & 9.24 & 0.193 & 0.153 & 0.224 & EvR+FF & \cite{zhang2011fusion,PhysRevC.60.044602} \\
$^{33}$S+$^{91}$Zr & 78.63 & 1.75 & 9.52 & 0 & 0.201 & 0.976 & EvR & \cite{corradi1990near} \\
$^{33}$S+$^{92}$Zr & 78.78 & 1.55 & 10.08 & 2.218 & 0.269 & 1.992 & EvR & \cite{corradi1990near} \\
$^{34}$S+$^{24}$Mg & 27.50 & 1.12 & 9.31 & 0.129 & 0.616 & 7.702 & EvR & \cite{PhysRevC.28.667} \\
$^{34}$S+$^{26}$Mg & 26.93 & 1.00 & 9.00 & 0 & 0.675 & 1.539 & EvR & \cite{PhysRevC.28.667} \\
$^{34}$S+$^{58}$Ni & 58.60 & 1.27 & 7.78 & 0 & 0.058 & 0.385 & EvR & \cite{STEFANINI1986509} \\
$^{34}$S+$^{64}$Ni & 56.42 & 1.37 & 8.52 & 0 & 0.412 & 0.250 & EvR & \cite{PhysRevC.42.1530} \\
$^{34}$S+$^{64}$Ni & 56.99 & 1.29 & 8.81 & 0 & 0.090 & 0.490 & EvR & \cite{STEFANINI1986509} \\
$^{34}$S+$^{89}$Y & 76.10 & 1.42 & 9.67 & 0 & 11.382 & 2.220 & EvR & \cite{PhysRevC.66.034607} \\
$^{34}$S+$^{168}$Er & 121.60 & 2.96 & 10.46 & 2.055 & 2.564 & 0.123 & EvR+FF & \cite{PhysRevC.62.024607} \\
$^{34}$S+$^{168}$Er & 121.44 & 2.88 & 10.36 & 2.109 & 0.065 & 0.045 & EvR+FF & \cite{MORTON2000160} \\
$^{34}$S+$^{168}$Er & 123.61 & 3.29 & 10.64 & 1.883 & 2.333 & 0.346 & FF & \cite{PhysRevC.62.024607} \\
$^{34}$S+$^{204}$Pb & 142.41 & 2.65 & 9.43 & 0 & 0.078 & 5.638 & FF & \cite{PhysRevC.86.064602} \\
$^{34}$S+$^{206}$Pb & 140.84 & 1.97 & 8.80 & 0 & 0.205 & 13.544 & FF & \cite{PhysRevC.86.064602} \\
$^{34}$S+$^{208}$Pb & 141.30 & 1.59 & 9.56 & 2.378 & 0.437 & 10.072 & FF & \cite{PhysRevC.86.064602} \\
$^{34}$S+$^{238}$U & 153.29 & 4.77 & 8.95 & 0 & 0.074 & 1.046 & FF & \cite{PhysRevC.82.024611} \\
$^{36}$S+$^{48}$Ca & 42.44 & 1.22 & 10.39 & 0.232 & 0.603 & 2.775 & EvR & \cite{PhysRevC.78.044607} \\
$^{36}$S+$^{48}$Ca & 42.15 & 1.09 & 11.13 & 0.238 & 1.091 & 21.585 & EvR & \cite{montagnoli2009fusion} \\
$^{36}$S+$^{58}$Ni & 58.30 & 1.42 & 7.58 & 0.171 & 0.105 & 0.644 & EvR & \cite{STEFANINI1986509} \\
$^{36}$S+$^{58}$Ni & 58.34 & 1.42 & 7.91 & 0.184 & 0.162 & 0.834 & EvR & \cite{STEFANINI198566} \\
$^{36}$S+$^{64}$Ni & 57.04 & 1.17 & 8.88 & 0.153 & 0.415 & 1.875 & EvR & \cite{STEFANINI198566} \\
$^{36}$S+$^{64}$Ni & 56.87 & 1.14 & 8.83 & 0.137 & 0.157 & 1.386 & EvR & \cite{STEFANINI1986509} \\
$^{36}$S+$^{64}$Ni & 56.25 & 1.19 & 9.79 & 0.121 & 5.629 & 2.783 & EvR & \cite{PhysRevC.82.064609} \\
$^{36}$S+$^{90}$Zr & 77.22 & 1.36 & 11.14 & 0 & 3.920 & 1.672 & EvR & \cite{PhysRevC.62.014601} \\
$^{36}$S+$^{96}$Zr & 75.35 & 1.00 & 11.46 & 3.648 & 3.854 & 4.935 & EvR & \cite{PhysRevC.62.014601} \\
$^{36}$S+$^{92}$Mo & 84.87 & 2.04 & 13.87 & 3.928 & 0.499 & 9.453 & EvR & \cite{PENGO1983255} \\
$^{36}$S+$^{94}$Mo & 80.82 & 1.91 & 10.25 & 0 & 0.665 & 7.829 & EvR & \cite{PENGO1983255} \\
$^{36}$S+$^{96}$Mo & 79.83 & 1.14 & 8.41 & 2.051 & 2.374 & 8.993 & EvR & \cite{PENGO1983255} \\
$^{36}$S+$^{98}$Mo & 78.61 & 0.83 & 8.68 & 2.667 & 0.810 & 1.136 & EvR & \cite{PENGO1983255} \\
$^{36}$S+$^{100}$Mo & 78.92 & 1.08 & 10.02 & 2.689 & 0.492 & 1.379 & EvR & \cite{PENGO1983255} \\
$^{36}$S+$^{100}$Ru & 82.93 & 1.24 & 8.45 & 1.542 & 0.731 & 14.787 & EvR & \cite{PENGO1983255} \\
$^{36}$S+$^{101}$Ru & 82.79 & 1.64 & 9.75 & 0 & 0.122 & 0.346 & EvR & \cite{PENGO1983255} \\
$^{36}$S+$^{102}$Ru & 81.97 & 1.41 & 8.42 & 0.021 & 7.317 & 3.355 & EvR & \cite{PENGO1983255} \\
$^{36}$S+$^{104}$Ru & 83.18 & 1.50 & 10.54 & 2.216 & 2.312 & 6.163 & EvR & \cite{PENGO1983255} \\
$^{36}$S+$^{106}$Pd & 87.76 & 1.48 & 11.02 & 3.750 & 0.955 & 4.883 & EvR & \cite{PENGO1983255} \\
$^{36}$S+$^{108}$Pd & 87.49 & 1.55 & 10.41 & 3.395 & 0.071 & 2.187 & EvR & \cite{PENGO1983255} \\
$^{36}$S+$^{110}$Pd & 86.08 & 1.52 & 8.57 & 1.819 & 7.026 & 0.932 & EvR & \cite{PhysRevC.52.R1727} \\
$^{36}$S+$^{110}$Pd & 84.74 & 1.53 & 7.96 & 0.161 & 1.484 & 19.173 & EvR & \cite{PENGO1983255} \\
$^{36}$S+$^{204}$Pb & 141.08 & 1.01 & 9.60 & 3.362 & 0.203 & 3.384 & FF & \cite{PhysRevC.86.064602} \\
$^{36}$S+$^{206}$Pb & 139.98 & 1.38 & 8.97 & 0 & 0.463 & 9.838 & FF & \cite{PhysRevC.86.064602} \\
$^{36}$S+$^{208}$Pb & 139.90 & 1.36 & 9.92 & 0 & 0.282 & 6.264 & FF & \cite{PhysRevC.86.064602} \\
$^{36}$S+$^{238}$U & 154.81 & 4.19 & 12.26 & 0.904 & 0.424 & 2.015 & FF & \cite{itkis2010fusion} \\
$^{35}$Cl+$^{24}$Mg & 30.22 & 1.65 & 9.84 & 0.192 & 1.597 & 1.928 & EvR & \cite{CAVALLARO1990174} \\
$^{35}$Cl+$^{25}$Mg & 30.08 & 1.85 & 9.31 & 0 & 0.980 & 1.047 & EvR & \cite{CAVALLARO1990174} \\
$^{35}$Cl+$^{26}$Mg & 29.47 & 1.93 & 8.29 & 0 & 0.378 & 0.370 & EvR & \cite{CAVALLARO1990174} \\
$^{35}$Cl+$^{27}$Al & 30.55 & 0.76 & 8.26 & 0 & 0.234 & 0.110 & EvR & \cite{PhysRevC.14.1808} \\
$^{35}$Cl+$^{51}$V & 51.66 & 1.59 & 10.38 & 0.012 & 0.153 & 0.190 & EvR & \cite{PhysRevC.41.2164} \\
$^{35}$Cl+$^{58}$Ni & 61.32 & 1.40 & 9.00 & 0 & 0.907 & 1.373 & EvR & \cite{PhysRevC.14.1808} \\
$^{35}$Cl+$^{60}$Ni & 61.03 & 2.12 & 9.27 & 0 & 0.511 & 1.583 & EvR & \cite{PhysRevC.14.1808} \\
$^{35}$Cl+$^{62}$Ni & 60.59 & 1.53 & 9.48 & 1.615 & 0.214 & 0.107 & EvR+FF & \cite{PhysRevC.14.1808} \\
$^{35}$Cl+$^{62}$Ni & 60.73 & 1.57 & 9.65 & 1.850 & 0.190 & 0.143 & EvR & \cite{PhysRevC.11.1701} \\
$^{35}$Cl+$^{62}$Ni & 60.71 & 1.69 & 9.59 & 1.318 & 0.089 & 0.067 & EvR & \cite{PhysRevC.14.1808} \\
$^{35}$Cl+$^{64}$Ni & 60.30 & 2.26 & 9.67 & 0 & 0.373 & 0.388 & EvR & \cite{PhysRevC.14.1808} \\
$^{35}$Cl+$^{92}$Zr & 82.50 & 2.33 & 9.74 & 0 & 2.757 & 1.044 & EvR & \cite{PhysRevC.64.064608} \\
$^{35}$Cl+$^{130}$Te & 102.37 & 2.71 & 11.67 & 0 & 0.191 & 0.281 & EvR+FF & \cite{PhysRevC.102.024615} \\
$^{37}$Cl+$^{24}$Mg & 29.39 & 1.98 & 8.54 & 0 & 0.467 & 0.493 & EvR & \cite{CAVALLARO1990174} \\
$^{37}$Cl+$^{25}$Mg & 28.86 & 0.91 & 8.34 & 1.775 & 2.435 & 4.320 & EvR & \cite{CAVALLARO1990174} \\
$^{37}$Cl+$^{26}$Mg & 28.61 & 0.87 & 7.59 & 1.779 & 0.452 & 0.441 & EvR & \cite{CAVALLARO1990174} \\
$^{37}$Cl+$^{59}$Co & 58.36 & 1.48 & 8.74 & 0 & 1.252 & 0.701 & EvR & \cite{DASGUPTA1992351} \\
$^{37}$Cl+$^{70}$Ge & 66.97 & 1.68 & 8.31 & 0 & 0.741 & 0.600 & EvR & \cite{PhysRevC.63.054611} \\
$^{37}$Cl+$^{72}$Ge & 67.01 & 1.60 & 8.90 & 0 & 1.121 & 0.300 & EvR & \cite{PhysRevC.63.054611} \\
$^{37}$Cl+$^{73}$Ge & 67.45 & 2.41 & 8.37 & 0 & 0.877 & 0.327 & EvR & \cite{PhysRevC.63.054611} \\
$^{37}$Cl+$^{74}$Ge & 68.04 & 2.52 & 9.75 & 0 & 0.361 & 0.175 & EvR & \cite{PhysRevC.63.054611} \\
$^{37}$Cl+$^{76}$Ge & 68.13 & 1.62 & 10.11 & 2.461 & 0.405 & 0.122 & EvR & \cite{PhysRevC.63.054611} \\
$^{37}$Cl+$^{98}$Mo & 84.98 & 1.93 & 8.38 & 0 & 0.127 & 0.366 & EvR & \cite{mahon1997fusion} \\
$^{40}$Ar+$^{110}$Pd & 97.26 & 4.08 & 11.04 & 0 & 0.457 & 1.256 & EvR & \cite{jahnke1982global} \\
$^{40}$Ar+$^{112}$Sn & 104.77 & 1.86 & 9.43 & 1.910 & 0.776 & 1.075 & EvR+FF & \cite{REISDORF1985212} \\
$^{40}$Ar+$^{116}$Sn & 104.89 & 1.99 & 9.65 & 2.267 & 0.371 & 0.194 & EvR+FF & \cite{REISDORF1985212} \\
$^{40}$Ar+$^{122}$Sn & 104.55 & 2.13 & 10.38 & 2.020 & 1.832 & 2.154 & EvR+FF & \cite{REISDORF1985212} \\
$^{40}$Ar+$^{144}$Sm & 125.75 & 2.01 & 9.24 & 1.674 & 2.289 & 2.525 & EvR+FF & \cite{REISDORF1985212} \\
$^{40}$Ar+$^{144}$Sm & 126.73 & 2.20 & 9.61 & 1.539 & 0.602 & 2.438 & EvR & \cite{stokstad1980sub} \\
$^{40}$Ar+$^{148}$Sm & 127.50 & 3.07 & 10.42 & 1.902 & 2.628 & 6.590 & EvR+FF & \cite{REISDORF1985212} \\
$^{40}$Ar+$^{148}$Sm & 127.36 & 3.01 & 10.11 & 2.778 & 0.766 & 3.722 & EvR & \cite{stokstad1980sub} \\
$^{40}$Ar+$^{154}$Sm & 125.41 & 3.66 & 9.71 & 1.981 & 8.926 & 16.085 & EvR+FF & \cite{REISDORF1985212} \\
$^{40}$Ar+$^{154}$Sm & 126.04 & 5.14 & 12.49 & 0 & 0.496 & 1.687 & EvR & \cite{stokstad1980sub} \\
$^{40}$Ar+$^{176}$Hf & 146.00 & 7.23 & 10.89 & 0 & 0.887 & 0.244 & FF & \cite{CLERC1984571} \\
$^{40}$Ar+$^{179}$Hf & 144.96 & 7.24 & 10.86 & 0 & 0.595 & 0.146 & FF & \cite{CLERC1984571} \\
$^{40}$Ar+$^{208}$Pb & 158.75 & 3.86 & 10.56 & 0 & 4.271 & 1.273 & FF & \cite{CLERC1984571} \\
$^{40}$Ca+$^{40}$Ca & 53.32 & 1.37 & 9.93 & 0.137 & 1.689 & 8.400 & EvR & \cite{PhysRevC.30.1223} \\
$^{40}$Ca+$^{40}$Ca & 53.51 & 1.31 & 10.15 & 0.165 & 0.636 & 7.953 & EvR & \cite{PhysRevC.85.024607} \\
$^{40}$Ca+$^{44}$Ca & 52.08 & 1.16 & 8.75 & 1.658 & 0.281 & 4.082 & EvR & \cite{PhysRevC.30.1223} \\
$^{40}$Ca+$^{48}$Ca & 51.28 & 1.47 & 7.86 & 0 & 0.328 & 2.327 & EvR & \cite{PhysRevC.30.1223} \\
$^{40}$Ca+$^{48}$Ca & 51.87 & 1.37 & 8.24 & 0.813 & 4.253 & 2.502 & EvR & \cite{PhysRevC.82.041601} \\
$^{40}$Ca+$^{48}$Ca & 51.68 & 1.67 & 11.29 & 0.043 & 8.322 & 1.094 & EvR & \cite{PhysRevC.65.011601} \\
$^{40}$Ca+$^{46}$Ti & 57.21 & 1.41 & 9.37 & 0 & 0.128 & 0.459 & EvR & \cite{R_Vandenbosch_1997} \\
$^{40}$Ca+$^{48}$Ti & 56.97 & 1.45 & 9.20 & 0 & 0.052 & 0.250 & EvR & \cite{R_Vandenbosch_1997} \\
$^{40}$Ca+$^{50}$Ti & 57.00 & 1.62 & 9.03 & 0 & 0.024 & 0.093 & EvR & \cite{R_Vandenbosch_1997} \\
$^{40}$Ca+$^{58}$Ni & 71.71 & 1.90 & 8.87 & 0 & 0.458 & 1.120 & EvR & \cite{PhysRevC.20.2219} \\
$^{40}$Ca+$^{58}$Ni & 71.15 & 1.30 & 8.60 & 0 & 0.342 & 0.890 & EvR & \cite{PhysRevC.90.044610} \\
$^{40}$Ca+$^{60}$Ni & 70.93 & 1.99 & 9.46 & 0 & 0.068 & 0.358 & EvR & \cite{PhysRevC.20.2219} \\
$^{40}$Ca+$^{62}$Ni & 70.66 & 2.32 & 9.39 & 0 & 0.257 & 0.472 & EvR+FF & \cite{PhysRevC.20.2219} \\
$^{40}$Ca+$^{62}$Ni & 70.80 & 2.39 & 9.52 & 0 & 0.341 & 0.668 & EvR & \cite{PhysRevC.20.2219} \\
$^{40}$Ca+$^{64}$Ni & 69.44 & 1.39 & 8.68 & 1.412 & 0.808 & 6.110 & EvR & \cite{PhysRevC.90.044610} \\
$^{40}$Ca+$^{90}$Zr & 96.18 & 1.58 & 10.00 & 0 & 5.326 & 0.226 & EvR & \cite{TIMMERS1998421} \\
$^{40}$Ca+$^{90}$Zr & 96.40 & 1.54 & 9.97 & 0 & 0.304 & 0.218 & EvR & \cite{TIMMERS199735} \\
$^{40}$Ca+$^{94}$Zr & 94.95 & 2.60 & 9.96 & 0.394 & 0.601 & 0.197 & EvR & \cite{PhysRevC.76.014610} \\
$^{40}$Ca+$^{96}$Zr & 94.09 & 2.16 & 9.62 & 1.443 & 3.419 & 0.115 & EvR & \cite{TIMMERS1998421} \\
$^{40}$Ca+$^{96}$Zr & 94.32 & 2.17 & 9.60 & 1.411 & 0.079 & 0.063 & EvR & \cite{TIMMERS199735} \\
$^{40}$Ca+$^{96}$Zr & 95.05 & 2.28 & 10.29 & 2.304 & 0.108 & 0.621 & EvR & \cite{STEFANINI2014639} \\
$^{40}$Ca+$^{124}$Sn & 113.22 & 2.19 & 9.62 & 0.869 & 2.543 & 0.248 & EvR & \cite{A_M_Stefanini_1997} \\
$^{40}$Ca+$^{124}$Sn & 113.34 & 2.26 & 9.56 & 0.792 & 1.399 & 0.119 & EvR & \cite{SCARLASSARA200099} \\
$^{40}$Ca+$^{192}$Os & 166.75 & 4.69 & 10.24 & 0.185 & 0.368 & 2.358 & FF & \cite{PhysRevC.54.3068} \\
$^{40}$Ca+$^{194}$Pt & 171.46 & 3.31 & 9.71 & 1.473 & 0.386 & 0.889 & FF & \cite{PhysRevC.54.3068} \\
$^{40}$Ca+$^{197}$Au & 174.59 & 6.84 & 10.61 & 0.179 & 3.159 & 1.645 & FF & \cite{PhysRevC.45.2861} \\
$^{40}$Ca+$^{208}$Pb & 176.68 & 3.91 & 9.96 & 2.503 & 5.646 & 0.098 & FF & \cite{PhysRevC.45.2861} \\
$^{40}$Ca+$^{238}$U & 192.86 & 5.60 & 8.57 & 0 & 0.182 & 0.464 & FF & \cite{PhysRevC.86.034608} \\
$^{48}$Ca+$^{48}$Ca & 51.51 & 1.10 & 10.43 & 0.202 & 2.655 & 36.945 & EvR & \cite{STEFANINI200995} \\
$^{48}$Ca+$^{48}$Ca & 51.04 & 0.88 & 10.68 & 0.814 & 3.075 & 0.798 & EvR & \cite{PhysRevC.65.011601} \\
$^{48}$Ca+$^{90}$Zr & 94.70 & 1.78 & 9.86 & 0 & 5.126 & 2.074 & EvR & \cite{PhysRevC.73.034606} \\
$^{48}$Ca+$^{96}$Zr & 93.43 & 1.31 & 10.06 & 3.871 & 7.090 & 0.737 & EvR & \cite{PhysRevC.73.034606} \\
$^{48}$Ca+$^{154}$Sm & 140.22 & 4.16 & 10.63 & 1.811 & 1.856 & 3.485 & EvR+FF & \cite{TROTTA2004245} \\
$^{48}$Ca+$^{154}$Sm & 138.49 & 3.44 & 10.65 & 2.228 & 2.308 & 3.587 & EvR+FF & \cite{stefanini2005fusion} \\
$^{48}$Ca+$^{208}$Pb & 175.48 & 2.89 & 12.53 & 0 & 2.821 & 18.852 & FF & \cite{PROKHOROVA200845} \\
$^{48}$Ca+$^{238}$U & 193.27 & 4.62 & 11.55 & 0.204 & 0.114 & 0.865 & FF & \cite{PhysRevC.86.034608} \\
$^{45}$Sc+$^{51}$V & 61.82 & 1.40 & 8.74 & 1.902 & 0.130 & 0.374 & EvR & \cite{DASGUPTA1992351} \\
$^{46}$Ti+$^{46}$Ti & 63.09 & 1.55 & 10.74 & 2.091 & 5.638 & 0.937 & EvR & \cite{PhysRevC.65.034609} \\
$^{46}$Ti+$^{64}$Ni & 77.05 & 2.03 & 9.58 & 1.464 & 1.147 & 0.442 & EvR & \cite{prasad1996study} \\
$^{46}$Ti+$^{90}$Zr & 106.23 & 2.37 & 10.45 & 1.007 & 0.343 & 0.237 & EvR & \cite{PhysRevC.41.1584} \\
$^{46}$Ti+$^{93}$Nb & 108.70 & 3.61 & 10.78 & 0 & 0.404 & 0.132 & EvR & \cite{PhysRevC.41.1584} \\
$^{48}$Ti+$^{58}$Fe & 71.50 & 1.43 & 8.58 & 1.425 & 0.226 & 2.531 & EvR & \cite{PhysRevC.92.064607} \\
$^{48}$Ti+$^{58}$Ni & 78.88 & 2.24 & 9.85 & 0.176 & 1.667 & 0.582 & EvR & \cite{PhysRevC.53.803} \\
$^{48}$Ti+$^{60}$Ni & 77.39 & 1.80 & 9.89 & 0.166 & 1.683 & 0.616 & EvR & \cite{PhysRevC.53.803} \\
$^{48}$Ti+$^{64}$Ni & 77.99 & 1.62 & 11.65 & 3.759 & 6.520 & 2.380 & EvR & \cite{PhysRevC.53.803} \\
$^{48}$Ti+$^{122}$Sn & 126.45 & 2.40 & 9.43 & 1.866 & 2.875 & 0.952 & EvR+FF & \cite{PhysRevC.51.1336} \\
$^{50}$Ti+$^{60}$Ni & 77.27 & 1.99 & 9.87 & 0.128 & 2.075 & 1.297 & EvR & \cite{prasad1996study} \\
$^{50}$Ti+$^{90}$Zr & 104.29 & 1.37 & 10.05 & 1.753 & 0.464 & 0.130 & EvR & \cite{PhysRevC.41.1584} \\
$^{50}$Ti+$^{93}$Nb & 106.76 & 1.81 & 10.16 & 1.788 & 0.438 & 0.580 & EvR & \cite{PhysRevC.41.1584} \\
$^{58}$Ni+$^{54}$Fe & 91.67 & 1.25 & 9.02 & 2.571 & 0.413 & 2.641 & EvR & \cite{PhysRevC.81.037601} \\
$^{58}$Ni+$^{54}$Fe & 91.90 & 1.33 & 9.41 & 2.414 & 5.177 & 2.836 & EvR & \cite{PhysRevC.82.014614} \\
$^{58}$Ni+$^{54}$Fe & 91.78 & 1.30 & 8.94 & 2.330 & 0.330 & 2.227 & EvR & \cite{PhysRevC.92.064607} \\
$^{58}$Ni+$^{58}$Ni & 99.19 & 1.47 & 8.09 & 2.535 & 1.094 & 3.762 & EvR & \cite{PhysRevC.23.1581} \\
$^{58}$Ni+$^{60}$Ni & 97.99 & 1.68 & 8.34 & 2.356 & 6.947 & 1.402 & EvR & \cite{PhysRevLett.74.864} \\
$^{58}$Ni+$^{64}$Ni & 95.43 & 1.98 & 8.49 & 0.397 & 0.392 & 1.773 & EvR & \cite{ACKERMANN199691} \\
$^{58}$Ni+$^{64}$Ni & 96.67 & 1.97 & 8.45 & 2.487 & 0.750 & 2.661 & EvR & \cite{PhysRevLett.45.1472} \\
$^{58}$Ni+$^{74}$Ge & 108.30 & 2.62 & 7.94 & 1.336 & 0.658 & 2.104 & EvR & \cite{PhysRevC.25.837} \\
$^{58}$Ni+$^{90}$Zr & 133.74 & 2.18 & 7.76 & 1.765 & 0.031 & 0.177 & EvR & \cite{scarlassara1991fusion} \\
$^{58}$Ni+$^{92}$Mo & 138.89 & 2.55 & 8.33 & 0 & 0.109 & 0.300 & EvR & \cite{REHM199331} \\
$^{58}$Ni+$^{100}$Mo & 139.59 & 3.50 & 10.09 & 2.490 & 0.084 & 0.235 & EvR & \cite{REHM199331} \\
$^{58}$Ni+$^{112}$Sn & 159.52 & 4.67 & 7.53 & 0.613 & 0.067 & 0.128 & EvR+FF & \cite{PhysRevC.36.1379} \\
$^{58}$Ni+$^{112}$Sn & 165.50 & 7.01 & 7.82 & 0.466 & 0.035 & 0.138 & EvR+FF & \cite{PhysRevC.34.2155} \\
$^{58}$Ni+$^{114}$Sn & 162.31 & 6.13 & 7.33 & 0 & 0.866 & 3.934 & EvR+FF & \cite{PhysRevC.34.2155} \\
$^{58}$Ni+$^{116}$Sn & 166.79 & 5.35 & 9.12 & 2.330 & 0.133 & 0.703 & EvR+FF & \cite{PhysRevC.34.2155} \\
$^{58}$Ni+$^{118}$Sn & 163.22 & 3.19 & 8.15 & 2.835 & 0.038 & 0.134 & EvR+FF & \cite{PhysRevC.34.2155} \\
$^{58}$Ni+$^{120}$Sn & 163.71 & 3.91 & 8.71 & 2.083 & 0.044 & 0.221 & EvR+FF & \cite{PhysRevC.34.2155} \\
$^{58}$Ni+$^{124}$Sn & 162.02 & 5.63 & 8.93 & 0 & 0.325 & 1.596 & EvR+FF & \cite{PhysRevC.34.2155} \\
$^{58}$Ni+$^{124}$Sn & 157.46 & 3.99 & 9.44 & 0 & 0.001 & 0.001 & EvR+FF & \cite{PhysRevC.36.1379} \\
$^{58}$Ni+$^{124}$Sn & 158.28 & 3.09 & 9.68 & 1.235 & 0.498 & 4.938 & EvR & \cite{PhysRevC.91.044602} \\
$^{60}$Ni+$^{89}$Y & 129.25 & 1.58 & 7.45 & 1.597 & 0.529 & 4.631 & EvR & \cite{PhysRevLett.89.052701} \\
$^{60}$Ni+$^{100}$Mo & 136.03 & 2.97 & 8.33 & 1.627 & 0.532 & 1.381 & EvR & \cite{stefanini2013fusion} \\
$^{64}$Ni+$^{58}$Ni & 96.48 & 3.03 & 7.91 & 0 & 0.037 & 0.037 & EvR & \cite{PhysRevC.25.837} \\
$^{64}$Ni+$^{64}$Ni & 94.99 & 1.47 & 10.16 & 1.699 & 0.515 & 2.542 & EvR & \cite{ACKERMANN199691} \\
$^{64}$Ni+$^{64}$Ni & 93.74 & 1.30 & 8.57 & 2.605 & 0.419 & 1.110 & EvR & \cite{PhysRevC.25.837} \\
$^{64}$Ni+$^{64}$Ni & 94.45 & 1.23 & 9.69 & 1.908 & 1.919 & 7.259 & EvR & \cite{ACKERMANN1995129} \\
$^{64}$Ni+$^{64}$Ni & 92.43 & 1.32 & 8.40 & 0.217 & 0.333 & 1.269 & EvR & \cite{PhysRevLett.93.012701} \\
$^{64}$Ni+$^{74}$Ge & 104.61 & 1.79 & 7.50 & 1.553 & 0.856 & 2.461 & EvR & \cite{PhysRevC.25.837} \\
$^{64}$Ni+$^{92}$Zr & 131.72 & 3.09 & 7.39 & 0.167 & 0.169 & 0.605 & EvR & \cite{STEFANINI1992453} \\
$^{64}$Ni+$^{96}$Zr & 128.96 & 2.39 & 8.09 & 1.763 & 0.173 & 0.444 & EvR & \cite{STEFANINI1992453} \\
$^{64}$Ni+$^{92}$Mo & 135.86 & 3.04 & 9.41 & 0 & 0.400 & 1.219 & EvR & \cite{REHM199331} \\
$^{64}$Ni+$^{100}$Mo & 134.56 & 3.34 & 7.78 & 0.793 & 0.246 & 0.979 & EvR & \cite{REHM199331} \\
$^{64}$Ni+$^{112}$Sn & 157.61 & 2.69 & 8.64 & 3.062 & 0.195 & 0.913 & EvR+FF & \cite{PhysRevC.34.2155} \\
$^{64}$Ni+$^{114}$Sn & 158.85 & 4.08 & 9.37 & 0 & 0.000 & 0.000 & EvR+FF & \cite{PhysRevC.34.2155} \\
$^{64}$Ni+$^{116}$Sn & 158.47 & 3.92 & 10.02 & 0 & 0.180 & 0.843 & EvR+FF & \cite{PhysRevC.34.2155} \\
$^{64}$Ni+$^{118}$Sn & 157.41 & 4.31 & 9.36 & 0 & 0.481 & 0.757 & EvR+FF & \cite{PhysRevLett.50.1563,PhysRevLett.55.803} \\
$^{64}$Ni+$^{120}$Sn & 156.44 & 3.54 & 9.33 & 0 & 0.036 & 0.178 & EvR+FF & \cite{PhysRevC.34.2155} \\
$^{64}$Ni+$^{122}$Sn & 155.99 & 2.26 & 9.41 & 3.600 & 0.018 & 0.083 & EvR+FF & \cite{PhysRevC.34.2155} \\
$^{64}$Ni+$^{124}$Sn & 154.87 & 2.85 & 9.60 & 0 & 0.034 & 0.172 & EvR+FF & \cite{PhysRevC.34.2155} \\
$^{74}$Ge+$^{74}$Ge & 122.45 & 3.47 & 8.21 & 1.515 & 0.585 & 1.154 & EvR & \cite{PhysRevC.28.1963} \\
$^{81}$Br+$^{94}$Zr & 156.41 & 4.39 & 7.58 & 0 & 0.011 & 0.044 & EvR & \cite{PhysRevC.29.1938} \\
$^{86}$Kr+$^{70}$Ge & 133.71 & 2.88 & 9.10 & 1.737 & 7.133 & 7.799 & EvR & \cite{REISDORF1985154} \\
$^{86}$Kr+$^{76}$Ge & 131.48 & 2.43 & 9.27 & 2.101 & 1.960 & 0.587 & EvR & \cite{REISDORF1985154} \\
$^{124}$Sn+$^{40}$Ca & 113.18 & 2.38 & 9.66 & 0.889 & 0.582 & 0.362 & EvR & \cite{PhysRevC.85.054603} \\
$^{124}$Sn+$^{48}$Ca & 113.69 & 1.80 & 9.99 & 1.273 & 2.509 & 1.601 & EvR & \cite{PhysRevC.85.054603} \\
$^{132}$Sn+$^{40}$Ca & 115.24 & 2.98 & 11.06 & 1.241 & 1.235 & 2.956 & EvR & \cite{PhysRevC.85.054603} \\
$^{132}$Sn+$^{58}$Ni & 159.45 & 4.26 & 11.32 & 0.708 & 0.152 & 0.199 & EvR+FF & \cite{PhysRevLett.107.202701} \\
$^{132}$Sn+$^{64}$Ni & 151.09 & 5.83 & 8.23 & 0 & 0.221 & 0.418 & EvR & \cite{liang2004enhanced} \\
$^{132}$Sn+$^{64}$Ni & 157.61 & 3.08 & 11.97 & 3.865 & 0.090 & 0.198 & EvR+FF & \cite{PhysRevC.75.054607} \\
$^{134}$Te+$^{40}$Ca & 116.79 & 2.28 & 8.79 & 0 & 0.531 & 0.426 & EvR & \cite{PhysRevC.87.064612} \\
$^{134}$Te+$^{64}$Ni & 164.21 & 2.67 & 9.22 & 1.816 & 0.317 & 0.455 & EvR & \cite{shapira2005measurement} \\
$^{208}$Pb+$^{50}$Ti & 194.55 & 2.22 & 7.83 & 2.791 & 0.797 & 0.675 & FF & \cite{BOCK1982334} \\

\hline\\
\end{longtable}

\newpage
\renewcommand{\arraystretch}{1.0}
\footnotesize

\end{appendix}


\begin{thebibliography}{100}

\bibitem{RevModPhys.72.733}
S.~Hofmann and G.~M\"unzenberg,
\newblock Rev. Mod. Phys. 72 (2000) 733.

\bibitem{PhysRevLett.104.142502}
Y.~T. Oganessian, F.~S. Abdullin, P.~D. Bailey, et al.,
\newblock Phys. Rev. Lett. 104 (2010) 142502.

\bibitem{PhysRevC.78.034610}
V.~Zagrebaev and W.~Greiner.
\newblock Phys. Rev. C 78 (2008) 034610.

\bibitem{adamian2000dynamical}
G.~Adamian, N.~Antonenko, A.~Diaz-Torres, et al.,
\newblock Nucl. Phys. A 671 (2000) 233.

\bibitem{PhysRevC.84.061601}
N.~Wang, J.~Tian and W.~Scheid.
\newblock Phys. Rev. C 84 (2011) 061601.

\bibitem{PhysRevC.84.064614}
V.~V. Sargsyan, G.~G. Adamian, N.~V. Antonenko, et al.,
\newblock Phys. Rev. C 84 (2011) 064614.

\bibitem{PhysRevC.62.044610}
W.~D. Myers and W.~J. \ifmmode \acute{S}\else
  \'{S}\fi{}wia\ifmmode~\mbox{\c{}}\else \c{}\fi{}tecki,
\newblock Phys. Rev. C 62 (2000) 044610.

\bibitem{dasgupta1998measuring}
M.~Dasgupta, D.~Hinde, N.~Rowley, et al.,
\newblock Annu. Rev. Nucl. Part. S 48 (1998) 401.

\bibitem{PhysRevC.92.064604}
T.~Ichikawa.
\newblock Phys. Rev. C 92 (2015) 064604.

\bibitem{PhysRevC.103.054601}
P.~W. Wen, C.~J. Lin, R.~G. Nazmitdinov, et al.,
\newblock Phys. Rev. C 103 (2021) 054601.

\bibitem{gamow1928quantentheorie}
G.~Gamow.
\newblock Z Phys. 51 (1928) 204.

\bibitem{PhysRevLett.31.766}
C.~Y. Wong.
\newblock Phys. Rev. Lett. 31 (1973) 766.

\bibitem{Denisov18}
V. Yu. Denisov and I.Yu. Sedykh, Eur. Phys. J. A 55 (2019) 153.

\bibitem{stelson1988neutron}
P.~Stelson.
\newblock Phys. Lett. B 205 (1988) 190.

\bibitem{PhysRevC.41.1584}
P.~H. Stelson, H.~J. Kim, M.~Beckerman, et al.,
\newblock Phys. Rev. C 41 (1990) 1584.

\bibitem{PhysRevC.69.024611}
K.~Siwek-Wilczy\ifmmode~\acute{n}\else \'{n}\fi{}ska and
  J.~Wilczy\ifmmode~\acute{n}\else \'{n}\fi{}ski.
\newblock Phys. Rev. C 69 (2004) 024611.

\bibitem{Jiang18}
C. L. Jiang, K. E. Rehm, B. B. Back,, et al., Eur. Phys. J. A 54 (2018) 218.

\bibitem{PhysRevC.105.034606}
P.~W. Wen, C.~J. Lin, H.~M. Jia, et al.,
\newblock Phys. Rev. C 105 (2022) 034606.

\bibitem{liu2006applications}
M.~Liu, N.~Wang, Z.~Li, et al.,
\newblock Nucl. Phys. A 768 (2006) 80.

\bibitem{wang2009heavy}
N.~Wang, M.~Liu and Y.~Yang,
\newblock Sci China Ser G-Phys Mech Astron 52 (2009) 1554.

\bibitem{PhysRevC.65.014607}
V.~I. Zagrebaev, Y.~Aritomo, M.~G. Itkis, et al.,
\newblock Phys. Rev. C 65 (2001) 014607.

\bibitem{WANG2017281}
B.~Wang, K.~Wen, W.-J. Zhao, et al.,
\newblock At. Data Nucl. Data Tables 114 (2017) 281.

\bibitem{wang2007systematic}
N.~Wang, Z.~Li and W.~Scheid,
\newblock J. Phys. G: Nucl. Part. Phys. 34 (2007) 1935.

\bibitem{PhysRevC.105.064601}
C.~L. Jiang and B.~P. Kay,
\newblock Phys. Rev. C 105 (2022) 064601.

\bibitem{ROWLEY199125}
N.~Rowley, G.~Satchler and P.~Stelson,
\newblock Phys. Lett. B 254 (1991) 25.

\bibitem{TIMMERS1998421}
H.~Timmers, D.~Ackermann, S.~Beghini, et al.,
\newblock Nucl. Phys. A 633 (1998) 421.

\bibitem{PhysRevC.17.126}
J.~R. Birkelund and J.~R. Huizenga.
\newblock Phys. Rev. C 17 (1978) 126.

\bibitem{PhysRevC.78.024610}
K.~Washiyama and D.~Lacroix.
\newblock Phys. Rev. C 78 (2008) 024610.

\bibitem{PhysRevC.105.024328}
H.~B. Zhou, Z.~Y. Li, Z.~G. Gan, et al.,
\newblock Phys. Rev. C 105 (2022) 024328.

\bibitem{PhysRevC.81.044602}
Y.~Jiang, N.~Wang, Z.~Li, et al.,
\newblock Phys. Rev. C 81 (2010) 044602.

\bibitem{PhysRevC.89.064601}
N.~Wang, L.~Ou, Y.~Zhang, et al.,
\newblock Phys. Rev. C 89 (2014) 064601.

\bibitem{PhysRevC.74.021601}
A.~S. Umar and V.~E. Oberacker,
\newblock Phys. Rev. C 74 (2006) 021601.

\bibitem{PhysRevC.80.041601}
A.~S. Umar, V.~E. Oberacker, J.~A. Maruhn, et al.,
\newblock Phys. Rev. C 80 (2009) 041601.

\bibitem{PhysRevC.60.044608}
C.~R. Morton, A.~C. Berriman, M.~Dasgupta, et al.,
\newblock Phys. Rev. C 60 (1999) 044608.

\bibitem{PhysRevC.62.024607}
C.~R. Morton, A.~C. Berriman, R.~D. Butt, et al.,
\newblock Phys. Rev. C 62 (2000) 024607.

\bibitem{PhysRevC.91.044617}
N.~Rowley and K.~Hagino,
\newblock Phys. Rev. C 91 (2015) 044617.

\bibitem{PhysRevC.82.054609}
H.~Q. Zhang, C.~J. Lin, F.~Yang, et al.,
\newblock Phys. Rev. C 82 (2010) 054609.

\bibitem{PhysRevC.52.R1727}
A.~M. Stefanini, D.~Ackermann, L.~Corradi, et al.,
\newblock Phys. Rev. C 52 (1995) R1727.

\bibitem{james1975minuit}
F.~James and M.~Roos,
\newblock Comput. Phys. Commun. 10 (1975) 343.

\bibitem{PhysRevC.82.014614}
A.~M. Stefanini, G.~Montagnoli, L.~Corradi, et al.,
\newblock Phys. Rev. C 82 (2010) 014614.

\bibitem{PhysRevLett.74.864}
A.~M. Stefanini, D.~Ackermann, L.~Corradi, et al.,
\newblock Phys. Rev. Lett. 74 (1995) 864.

\bibitem{nrvweb}
{http://nrv.jinr.ru/nrv/}.

\bibitem{SWIATECKI1982275}
W.~Swiatecki,
\newblock Nucl. Phys. A 376 (1982) 275.

\bibitem{PhysRevC.88.011301}
N.~Wang and T.~Li.
\newblock Phys. Rev. C 88 (2013) 011301.

\bibitem{Angeli13}
I. Angeli, K.P. Marinova, At. Data Nucl. Data Tables 99 (2013) 69.

\bibitem{Litao21}
T. Li, Y. Luo, N. Wang, At. Data Nucl. Data Tables 140 (2021) 101440.

\bibitem{Swiateck05}
W. J. \'Swi\c{a}tecki, K. Siwek-Wilczy\'nska, and J. Wilczy\'nski,
Phys. Rev. C 71 (2005) 014602.


\bibitem{PhysRevC.82.044608}
C.~S. Palshetkar, S.~Santra, A.~Chatterjee, et al.,
\newblock Phys. Rev. C 82 (2010) 044608.

\bibitem{PhysRevC.31.1752}
S.~Gil, R.~Vandenbosch, A.~J. Lazzarini, et al.,
\newblock Phys. Rev. C 31 (1985) 1752.

\bibitem{FREIESLEBEN1974503}
H.~Freiesleben and J.~Huizenga,
\newblock Nucl. Phys. A 224 (1974) 503.

\bibitem{PhysRev.115.1247}
R.~Gunnink and J.~W. Cobble,
\newblock Phys. Rev. 115 (1959) 1247.

\bibitem{LIMKILDE1973504}
P.~Limkilde and G.~Sletten,
\newblock Nucl. Phys. A 199 (1973) 504.

\bibitem{PhysRev.121.1415}
L.~J. Colby, M.~L. Shoaf and J.~W. Cobble,
\newblock Phys. Rev. 121 (1961) 1415.

\bibitem{PhysRevC.7.1231}
A.~Fleury, F.~H. Ruddy, M.~N. Namboodiri, et al.,
\newblock Phys. Rev. C 7 (1973) 1231.

\bibitem{fisichella2011halo}
M.~Fisichella, V.~Scuderi, A.~Di~Pietro, et al.,
\newblock J. Phys. Conf. Ser. 282 (2011) 012014.

\bibitem{PhysRevLett.81.4580}
J.~J. Kolata, V.~Guimar\~aes, D.~Peterson, et al.,
\newblock Phys. Rev. Lett. 81 (1998) 4580.

\bibitem{PhysRevLett.103.232701}
A.~Lemasson, A.~Shrivastava, A.~Navin, et al.,
\newblock Phys. Rev. Lett. 103 (2009) 232701.

\bibitem{PhysRevC.91.034615}
M.~M. Shaikh, S.~Roy, S.~Rajbanshi, et al.,
\newblock Phys. Rev. C 91 (2015) 034615.

\bibitem{PhysRevC.87.064614}
A.~Di~Pietro, P.~Figuera, E.~Strano, et al.,
\newblock Phys. Rev. C 87 (2013) 064614.

\bibitem{PhysRevC.86.024607}
H.~Kumawat, V.~Jha, V.~V. Parkar, et al.,
\newblock Phys. Rev. C 86 (2012) 024607.

\bibitem{PhysRevC.79.051601}
P.~K. Rath, S.~Santra, N.~L. Singh, et al.,
\newblock Phys. Rev. C 79 (2009) 051601.

\bibitem{RATH201214}
P.~Rath, S.~Santra, N.~Singh, et al.,
\newblock Nucl. Phys. A 874 (2012) 14.

\bibitem{PhysRevC.83.064606}
M.~K. Pradhan, A.~Mukherjee, P.~Basu, et al.,
\newblock Phys. Rev. C 83 (2011) 064606.

\bibitem{PhysRevLett.103.232702}
A.~Shrivastava, A.~Navin, A.~Lemasson, et al.,
\newblock Phys. Rev. Lett. 103 (2009) 232702.

\bibitem{PhysRevC.89.024607}
C.~S. Palshetkar, S.~Thakur, V.~Nanal, et al.,
\newblock Phys. Rev. C 89 (2014) 024607.

\bibitem{PhysRevC.66.041602}
M.~Dasgupta, D.~J. Hinde, K.~Hagino, et al.,
\newblock Phys. Rev. C 66 (2002) 041602.

\bibitem{PhysRevC.12.42}
H.~Freiesleben, G.~T. Rizzo and J.~R. Huizenga.
\newblock Phys. Rev. C 12 (1975) 42.

\bibitem{scholz1986complete}
C.~Scholz, L.~Ricken and E.~Kuhlmann,
\newblock Z. Phys. A 325 (1986) 203.

\bibitem{pakou2009total}
A.~Pakou, K.~Rusek, N.~Alamanos, et al.,
\newblock Eur. Phys. J. A 39 (2009) 187.

\bibitem{PhysRevC.67.054602}
C.~Beck, F.~A. Souza, N.~Rowley, et al.,
\newblock Phys. Rev. C 67 (2003) 054602.

\bibitem{PhysRevC.88.044617}
P.~K. Rath, S.~Santra, N.~L. Singh, et al.,
\newblock Phys. Rev. C 88 (2013) 044617.

\bibitem{BRODA1975356}
R.~Broda, M.~Ishihara, B.~Herskind, et al.,
\newblock Nucl. Phys. A 248 (1975) 356.

\bibitem{SHRIVASTAVA2013931}
A.~Shrivastava, A.~Navin, A.~Diaz-Torres, et al.,
\newblock Phys. Lett. B 718 (2013) 931.

\bibitem{PhysRevC.10.245}
H.~Freiesleben, H.~C. Britt, J.~Birkelund, et al.,
\newblock Phys. Rev. C 10 (1974) 245.

\bibitem{PhysRevC.90.014603}
A.~Parihari, S.~Santra, A.~Pal, et al.,
\newblock Phys. Rev. C 90 (2014) 014603.

\bibitem{PhysRevC.90.014616}
E.~Martinez-Quiroz, E.~F. Aguilera, D.~Lizcano, et al.,
\newblock Phys. Rev. C 90 (2014) 014616.

\bibitem{PhysRevC.74.044606}
R.~Raabe, C.~Angulo, J.~L. Charvet, et al.,
\newblock Phys. Rev. C 74 (2006) 044606.

\bibitem{PhysRevC.82.054601}
V.~V. Parkar, R.~Palit, S.~K. Sharma, et al.,
\newblock Phys. Rev. C 82 (2010) 054601.

\bibitem{PhysRevC.73.064606}
P.~R.~S. Gomes, I.~Padron, E.~Crema, et al.,
\newblock Phys. Rev. C 73 (2006) 064606.

\bibitem{PhysRevC.91.014608}
Y.~D. Fang, P.~R.~S. Gomes, J.~Lubian, et al.,
\newblock Phys. Rev. C 91 (2015) 014608.

\bibitem{PhysRevC.90.024621}
N.~T. Zhang, Y.~D. Fang, P.~R.~S. Gomes, et al.,
\newblock Phys. Rev. C 90 (2014) 024621.

\bibitem{PhysRevC.70.024606}
M.~Dasgupta, P.~R.~S. Gomes, D.~J. Hinde, et al.,
\newblock Phys. Rev. C 70 (2004) 024606.

\bibitem{liu2005partial}
Z.~Liu, C.~Signorini, M.~Mazzocco, et al.,
\newblock Eur. Phys. J. A 26 (2005) 73.

\bibitem{signorini1999does}
C.~Signorini, Z.~Liu, Z.~Li, et al.,
\newblock Eur. Phys. J. A 5 (1999) 7.

\bibitem{signorini1998fusion}
C.~Signorini, Z.~Liu, A.~Yoshida, et al.,
\newblock Eur. Phys. J. A 2 (1998) 227.

\bibitem{MUKHERJEE200691}
A.~Mukherjee, {Subinit Roy}, M.~Pradhan, et al.,
\newblock Phys. Lett. B 636 (2006) 91.

\bibitem{DASMAHAPATRA1983192}
B.~Dasmahapatra, B.~$\check{\rm C}$ujec and F.~Lahlou,
\newblock Nucl. Phys. A 408 (1983) 192.

\bibitem{liu1996fission}
Z.~Liu, H.~Zhang, J.~Xu, et al.,
\newblock Phys. Rev. C 54 (1996) 761.

\bibitem{CHEUNG1978333}
H.~Cheung, M.~High and B.~$\check{\rm C}$ujec,
\newblock Nucl. Phys. A 296 (1978) 333.

\bibitem{HIGH1977149}
M.~High and B.~$\check{\rm C}$ujec,
\newblock Nucl. Phys. A 278 (1977) 149.

\bibitem{PhysRevLett.37.888}
R.~G. Stokstad, Z.~E. Switkowski, R.~A. Dayras, et al.,
\newblock Phys. Rev. Lett. 37 (1976) 888.

\bibitem{DASMAHAPATRA1993657}
B.~Dasmahapatra and B.~$\check{\rm C}$ujec,
\newblock Nucl. Phys. A 565 (1993) 657.

\bibitem{hulke1980comparison}
G.~Hulke, C.~Rolfs and H.~Trautvetter.
\newblock Z. Phys. A 297 (1980) 161.

\bibitem{BOZEK1986171}
E.~Bozek, D.~{De Castro-Rizzo}, S.~Cavallaro, et al.,
\newblock Nucl. Phys. A 451 (1986) 171.

\bibitem{DUMONT1985301}
H.~Dumont, B.~Delaunay, J.~Delaunay, et al.,
\newblock Nucl. Phys. A 435 (1985) 301.

\bibitem{PhysRevC.64.064608}
J.~O. Newton, C.~R. Morton, M.~Dasgupta, et al.,
\newblock Phys. Rev. C 64 (2001) 064608.

\bibitem{PhysRevC.46.244}
D.~Abriola, A.~A. Sonzogni, M.~di~Tada, et al.,
\newblock Phys. Rev. C 46 (1992) 244.

\bibitem{crippa1994excitation}
M.~Crippa, E.~Gadioli, P.~Vergani, et al.,
\newblock Z. Phys. A 350 (1994) 121.

\bibitem{PhysRevC.63.054602}
A.~Shrivastava, S.~Kailas, A.~Chatterjee, et al.,
\newblock Phys. Rev. C 63 (2001) 054602.

\bibitem{PhysRevC.68.014603}
R.~N. Sagaidak, G.~N. Kniajeva, I.~M. Itkis, et al.,
\newblock Phys. Rev. C 68 (2003) 014603.

\bibitem{PhysRevC.75.044608}
A.~Mukherjee, D.~J. Hinde, M.~Dasgupta, et al.,
\newblock Phys. Rev. C 75 (2007) 044608.

\bibitem{CHATTERJEE1980273}
M.~L. Chatterjee, L.~Potvin and B.~$\check{\rm C}$ujec,
\newblock Nucl. Phys. A 333 (1980) 273.

\bibitem{PhysRevC.72.067601}
B.~P. Ajith~Kumar, K.~M. Varier, R.~G. Thomas, et al.,
\newblock Phys. Rev. C 72 (2005) 067601.

\bibitem{SHIUCHINWU1978177}
{Shiu-Chin Wu}, J.~Overley, C.~Barnes, et al.,
\newblock Nucl. Phys. A 312 (1978) 177.

\bibitem{SWITKOWSKI1977502}
Z.~Switkowski, R.~Stokstad and R.~Wieland,
\newblock Nucl. Phys. A 279 (1977) 502.

\bibitem{PhysRevC.26.1482}
P.~A. DeYoung, J.~J. Kolata, L.~J. Satkowiak, et al.,
\newblock Phys. Rev. C 26 (1982) 1482.

\bibitem{SWITKOWSKI1976202}
Z.~Switkowski, R.~Stokstad and R.~Wieland,
\newblock Nucl. Phys. A 274 (1976) 202.

\bibitem{GOMES1989395}
P.~Gomes, T.~Penna, R.~{Liguori Neto}, et al.,
\newblock Nucl. Instrum. Meth. A 280 (1989) 395.

\bibitem{PhysRevC.69.064603}
B.~R. Behera, M.~Satpathy, S.~Jena, et al.,
\newblock Phys. Rev. C 69 (2004) 064603.

\bibitem{FUNAKI1993307}
H.~Funaki and E.~Arai.
\newblock Nucl. Phys. A 556 (1993) 307.

\bibitem{PhysRevC.33.2017}
E.~Vulgaris, L.~Grodzins, S.~G. Steadman, et al.,
\newblock Phys. Rev. C 33 (1986) 2017.

\bibitem{CHRISTENSEN1977189}
P.~Christensen, Z.~Switkowskiw and R.~Dayras,
\newblock Nucl. Phys. A 280 (1977) 189.

\bibitem{DASMAHAPATRA1991395}
B.~Dasmahapatra, B.$\check{\rm C}$ujec, I.~Sz\"oghy, et al.,
\newblock Nucl. Phys. A 526 (1991) 395.

\bibitem{PhysRevC.35.591}
A.~Kuronen, J.~Keinonen and P.~Tikkanen.
\newblock Phys. Rev. C 35 (1987) 591.

\bibitem{PhysRevC.33.1679}
J.~Thomas, Y.~T. Chen, S.~Hinds, et al.,
\newblock Phys. Rev. C 33 (1986) 1679.

\bibitem{DAUK1975170}
J.~Dauk, K.~Lieb and A.~Kleinfeld,
\newblock Nucl. Phys. A 241 (1975) 170.

\bibitem{NETO1990333}
R.~L. Neto, J.~Acquadro, P.~Gomes, et al.,
\newblock Nucl. Phys. A 512 (1990) 333.

\bibitem{KEELEY19981}
N.~Keeley, J.~Lilley, J.~Wei, et al.,
\newblock Nucl. Phys. A 628 (1998) 1.

\bibitem{CHAMON199229}
L.~Chamon, D.~Pereira, E.~Rossi, et al.,
\newblock Phys. Lett. B 275 (1992) 29.

\bibitem{PEREIRA1989347}
D.~Pereira, G.~Ramirez, O.~Sala, et al.,
\newblock Phys. Lett. B 220 (1989) 347.

\bibitem{PhysRevC.14.152}
M.~Langevin, J.~Barreto and C.~D\'etraz,
\newblock Phys. Rev. C 14 (1976) 152.

\bibitem{PhysRevC.71.034608}
P.~R.~S. Gomes, M.~D. Rodr\'{\i}guez, G.~V. Mart\'{\i}, et al.,
\newblock Phys. Rev. C 71 (2005) 034608.

\bibitem{PhysRevC.52.3103}
E.~F. Aguilera, J.~J. Kolata and R.~J. Tighe,
\newblock Phys. Rev. C 52 (1995) 3103.

\bibitem{PhysRevC.86.044621}
H.~M. Jia, C.~J. Lin, F.~Yang, et al.,
\newblock Phys. Rev. C 86 (2012) 044621.

\bibitem{ACKERMANN1994374}
D.~Ackermann, L.~Corradi, D.~Napoli, et al.,
\newblock Nucl. Phys. A 575 (1994) 374.

\bibitem{PhysRevC.65.014614}
V.~Tripathi, L.~T. Baby, J.~J. Das, et al.,
\newblock Phys. Rev. C 65 (2001) 014614.

\bibitem{PhysRevC.47.2970}
M.~di~Tada, D.~E. DiGregorio, D.~Abriola, et al.,
\newblock Phys. Rev. C 47 (1993) 2970.

\bibitem{PhysRevC.52.3151}
J.~R. Leigh, M.~Dasgupta, D.~J. Hinde, et al.,
\newblock Phys. Rev. C 52 (1995) 3151.

\bibitem{PhysRevC.39.516}
D.~E. DiGregorio, M.~diTada, D.~Abriola, et al.,
\newblock Phys. Rev. C 39 (1989) 516.

\bibitem{PhysRevC.21.2427}
R.~G. Stokstad, Y.~Eisen, S.~Kaplanis, et al.,
\newblock Phys. Rev. C 21 (1980) 2427.

\bibitem{jahnke1982global}
U.~Jahnke, H.~Rossner, D.~Hilscher, et al.,
\newblock Phys. Rev. Lett. 48 (1982) 17.

\bibitem{PhysRevLett.67.3368}
J.~X. Wei, J.~R. Leigh, D.~J. Hinde, et al.,
\newblock Phys. Rev. Lett. 67 (1991) 3368.

\bibitem{PhysRevC.43.2303}
J.~O. Fern\'andez~Niello, M.~di~Tada, A.~O. Macchiavelli, et al.,
\newblock Phys. Rev. C 43 (1991) 2303.

\bibitem{PhysRevC.93.054622}
T.~Rajbongshi, K.~Kalita, S.~Nath, et al.,
\newblock Phys. Rev. C 93 (2016) 054622.

\bibitem{LEMMON199332}
R.~Lemmon, J.~Leigh, J.~Wei, et al.,
\newblock Phys. Lett. B 316 (1993) 32.

\bibitem{trotta2005fusion}
M.~Trotta, A.~Stefanini, S.~Beghini, et al.,
\newblock Eur. Phys. J. A 25 (2005) 615.

\bibitem{PRASAD201262}
E.~Prasad, K.~Varier, R.~Thomas, et al.,
\newblock Nucl. Phys. A 882 (2012) 62.

\bibitem{PhysRevC.84.064606}
E.~Prasad, K.~M. Varier, N.~Madhavan, et al.,
\newblock Phys. Rev. C 84 (2011) 064606.

\bibitem{HINDE1986550}
D.~Hinde, R.~Charity, G.~Foote, et al.,
\newblock Nucl. Phys. A 452 (1986) 550.

%\bibitem{PhysRevC.50.309}
%K.~T. Brinkmann, A.~L. Caraley, B.~J. Fineman, et al.,
%\newblock Phys. Rev. C 50 (1994) 309.

\bibitem{PhysRev.135.B669}
T.~Sikkeland,
\newblock Phys. Rev. 135 (1964) B669.

\bibitem{PhysRevC.13.1527}
Y.~Eyal, M.~Beckerman, R.~Chechik, et al.,
\newblock Phys. Rev. C 13 (1976) 1527.

\bibitem{PhysRevC.47.2699}
R.~J. Tighe, J.~J. Kolata, M.~Belbot, et al.,
\newblock Phys. Rev. C 47 (1993) 2699.

\bibitem{ROTH1980148}
H.~Roth, J.~Christiansson and J.~Dubois,
\newblock Nucl. Phys. A 343 (1980) 148.

\bibitem{PhysRevC.90.041603}
T.~K. Steinbach, J.~Vadas, J.~Schmidt, et al.,
\newblock Phys. Rev. C 90 (2014) 041603.

\bibitem{PhysRevC.46.2360}
A.~M. Borges, C.~P. da~Silva, D.~Pereira, et al.,
\newblock Phys. Rev. C 46 (1992) 2360.

\bibitem{PhysRevC.55.3155}
C.~P. Silva, D.~Pereira, L.~C. Chamon, et al.,
\newblock Phys. Rev. C 55 (1997) 3155.

\bibitem{prasad1996study}
N.~V. Prasad, A.~Vinodkumar, A.~Sinha, et al.,
\newblock Nucl. Phys. A 603 (1996).

\bibitem{HINDE1982109}
D.~Hinde, J.~Leigh, J.~Newton, et al.,
\newblock Nucl. Phys. A 385 (1982) 109.

\bibitem{MAHATA2003209}
K.~Mahata, S.~Kailas, A.~Shrivastava, et al.,
\newblock Nucl. Phys. A 720 (2003) 209.

\bibitem{PhysRevC.60.054602}
D.~J. Hinde, A.~C. Berriman, M.~Dasgupta, et al.,
\newblock Phys. Rev. C 60 (1999) 054602.

\bibitem{HUANQIAO1990531}
Z.~Huanqiao, L.~Zuhua, X.~Jincheng, et al.,
\newblock Nucl. Phys. A 512 (1990) 531.

\bibitem{PhysRevLett.81.3341}
K.~E. Rehm, H.~Esbensen, C.~L. Jiang, et al.,
\newblock Phys. Rev. Lett. 81 (1998) 3341.

\bibitem{samant2000fission}
A.~Samant, S.~Kailas, A.~Chatterjee, et al.,
\newblock Eur. Phys. J. A 7 (2000) 59.

\bibitem{PhysRevC.51.3109}
N.~Majumdar, P.~Bhattacharya, D.~C. Biswas, et al.,
\newblock Phys. Rev. C 51 (1995) 3109.

\bibitem{PhysRevC.43.1466}
S.~Kailas, A.~Navin, A.~Chatterjee, et al.,
\newblock Phys. Rev. C 43 (1991) 1466.

\bibitem{ZHANG1989133}
H.~Zhang, J.~Xu, Z.~Liu, et al.,
\newblock Phys. Lett. B 218 (1989) 133.

\bibitem{PhysRevC.85.054608}
E.~Piasecki, L.~\ifmmode~\acute{S}\else \'{S}\fi{}widerski, N.~Keeley, et al.,
\newblock Phys. Rev. C 85 (2012) 054608.

\bibitem{PhysRev.128.767}
V.~E. Viola and T.~Sikkeland,
\newblock Phys. Rev. 128 (1962) 767.

\bibitem{PhysRevLett.57.2002}
R.~Butsch, H.~J\"ansch, D.~Kr\"amer, et al.,
\newblock Phys. Rev. Lett. 57 (1986) 2002.

\bibitem{PhysRevC.25.1877}
S.~Gary and C.~Volant,
\newblock Phys. Rev. C 25 (1982) 1877.

\bibitem{PhysRevLett.113.022701}
C.~L. Jiang, A.~M. Stefanini, H.~Esbensen, et al.,
\newblock Phys. Rev. Lett. 113 (2014) 022701.

\bibitem{itkis2010fusion}
M.~Itkis, I.~Itkis, G.~Knyazheva, et al.,
\newblock Nucl. Phys. A 834 (2010) 374c.

\bibitem{PhysRevC.81.024611}
C.~L. Jiang, K.~E. Rehm, H.~Esbensen, et al.,
\newblock Phys. Rev. C 81 (2010) 024611.

\bibitem{PhysRevC.41.910}
E.~F. Aguilera, J.~J. Vega, J.~J. Kolata, et al.,
\newblock Phys. Rev. C 41 (1990) 910.

\bibitem{watanabe2001measurement}
Y.~Watanabe, A.~Yoshida, T.~Fukuda, et al.,
\newblock Eur. Phys. J. A 10 (2001) 373.

\bibitem{PhysRevC.41.988}
A.~Morsad, J.~J. Kolata, R.~J. Tighe, et al.,
\newblock Phys. Rev. C 41 (1990) 988.

\bibitem{PhysRevC.90.044608}
G.~Montagnoli, A.~M. Stefanini, H.~Esbensen, et al.,
\newblock Phys. Rev. C 90 (2014) 044608.

\bibitem{PhysRevC.78.017601}
C.~L. Jiang, B.~B. Back, H.~Esbensen, et al.,
\newblock Phys. Rev. C 78 (2008) 017601.

\bibitem{PhysRevC.30.2088}
A.~M. Stefanini, G.~Fortuna, A.~Tivelli, et al.,
\newblock Phys. Rev. C 30 (1984) 2088.

\bibitem{STEFANINI1986509}
A.~Stefanini, G.~Fortuna, R.~Pengo, et al.,
\newblock Nucl. Phys. A 456 (1986) 509.

\bibitem{JIANG200618}
C.~Jiang, B.~Back, H.~Esbensen, et al.,
\newblock Phys. Lett. B 640 (2006) 18.

\bibitem{A_K_Sinha_1997}
A.~K. Sinha, L.~T. Baby, N.~Badiger, et al.,
\newblock J. Phys. G: Nucl. Part. Phys. 23 (1997) 1331.

\bibitem{DASGUPTA1992351}
M.~Dasgupta, A.~Navin, Y.~K. Agarwal, et al.,
\newblock Nucl. Phys. A 539 (1992) 351.

\bibitem{PhysRevLett.66.1414}
M.~Dasgupta, A.~Navin, Y.~K. Agarwal, et al.,
\newblock Phys. Rev. Lett. 66 (1991) 1414.

\bibitem{PhysRevC.81.044610}
S.~Kalkal, S.~Mandal, N.~Madhavan, et al.,
\newblock Phys. Rev. C 81 (2010) 044610.

\bibitem{PhysRevC.96.014614}
Khushboo, S.~Mandal, S.~Nath, et al.,
\newblock Phys. Rev. C 96 (2017) 014614.

\bibitem{PhysRevC.56.1936}
L.~T. Baby, V.~Tripathi, D.~O. Kataria, et al.,
\newblock Phys. Rev. C 56 (1997) 1936.

\bibitem{ACKERMANN199691}
D.~Ackermann, P.~Bednarczyk, L.~Corradi, et al.,
\newblock Nucl. Phys. A 609 (1996) 91.

\bibitem{ACKERMANN1995129}
D.~Ackermann, F.~Scarlassara, P.~Bednarczyk, et al.,
\newblock Nucl. Phys. A 583 (1995) 129.

\bibitem{PhysRevLett.65.3100}
S.~Gil, D.~Abriola, D.~E. DiGregorio, et al.,
\newblock Phys. Rev. Lett. 65 (1990) 3100.

\bibitem{PhysRevC.66.044601}
R.~D. Butt, D.~J. Hinde, M.~Dasgupta, et al.,
\newblock Phys. Rev. C 66 (2002) 044601.

\bibitem{PhysRevC.62.014602}
K.~Nishio, H.~Ikezoe, S.~Mitsuoka, et al.,
\newblock Phys. Rev. C 62 (2000) 014602.

\bibitem{HINDE1995271}
D.~Hinde, C.~Morton, M.~Dasgupta, et al.,
\newblock Nucl. Phys. A 592 (1995) 271.

\bibitem{nishio2006measurement}
K.~Nishio, S.~Hofmann, F.~He{\ss}berger, et al.,
\newblock Eur. Phys. J. A 29 (2006) 281.

\bibitem{PhysRevC.82.044604}
K.~Nishio, H.~Ikezoe, I.~Nishinaka, et al.,
\newblock Phys. Rev. C 82 (2010) 044604.

\bibitem{PhysRevC.41.2654}
A.~Menchaca-Rocha, E.~Belmont-Moreno, M.~E. Brandan, et al.,
\newblock Phys. Rev. C 41 (1990) 2654.

\bibitem{PhysRevC.28.667}
G.~M. Berkowitz, P.~Braun-Munzinger, J.~S. Karp, et al.,
\newblock Phys. Rev. C 28 (1983) 667.

\bibitem{PhysRevC.87.014611}
G.~Montagnoli, A.~M. Stefanini, H.~Esbensen, et al.,
\newblock Phys. Rev. C 87 (2013) 014611.

\bibitem{PhysRevC.42.1530}
R.~J. Tighe, J.~J. Vega, E.~Aguilera, et al.,
\newblock Phys. Rev. C 42 (1990) 1530.

\bibitem{STEFANINI198566}
A.~Stefanini, G.~Fortuna, R.~Pengo, et al.,
\newblock Phys. Lett. B 162 (1985) 66.

\bibitem{PhysRevC.66.034607}
A.~Mukherjee, M.~Dasgupta, D.~J. Hinde, et al.,
\newblock Phys. Rev. C 66 (2002) 034607.

\bibitem{PhysRevC.89.064605}
H.~M. Jia, C.~J. Lin, F.~Yang, et al.,
\newblock Phys. Rev. C 89 (2014) 064605.

\bibitem{PENGO1983255}
R.~Pengo, D.~Evers, K.~L\"obner, et al.,
\newblock Nucl. Phys. A 411 (1983) 255.

\bibitem{PhysRevC.51.1336}
S.~Gil, F.~Hasenbalg, J.~E. Testoni, et al.,
\newblock Phys. Rev. C 51 (1995) 1336.

%\bibitem{PhysRevC.29.486}
%B.~G. Glagola, B.~B. Back and R.~R. Betts.
%\newblock Phys. Rev. C 29 (1984) 486.

\bibitem{PhysRevC.49.245}
P.~R.~S. Gomes, I.~C. Charret, R.~Wanis, et al.,
\newblock Phys. Rev. C 49 (1994) 245.

\bibitem{PhysRevC.62.054603}
S.~Mitsuoka, H.~Ikezoe, K.~Nishio, et al.,
\newblock Phys. Rev. C 62 (2000) 054603.

\bibitem{zhang2011fusion}
H.~Zhang, C.~Zhang, C.~Lin, et al.,
\newblock J. Phys. Conf. Ser. 282 (2011) 012013.

\bibitem{PhysRevC.60.044602}
B.~B. Back, D.~J. Blumenthal, C.~N. Davids, et al.,
\newblock Phys. Rev. C 60 (1999) 044602.

\bibitem{corradi1990near}
L.~Corradi, S.~Skorka, U.~Lenz, et al.,
\newblock Z. Phys. A 335 (1990) 55.

\bibitem{MORTON2000160}
C.~Morton, D.~Hinde, A.~Berriman, et al.,
\newblock Phys. Lett. B 481 (2000) 160.

\bibitem{PhysRevC.86.064602}
J.~Khuyagbaatar, K.~Nishio, S.~Hofmann, et al.,
\newblock Phys. Rev. C 86 (2012) 064602.

\bibitem{PhysRevC.82.024611}
K.~Nishio, S.~Hofmann, F.~P. He\ss{}berger, et al.,
\newblock Phys. Rev. C 82 (2010) 024611.

\bibitem{PhysRevC.78.044607}
A.~M. Stefanini, G.~Montagnoli, R.~Silvestri, et al.,
\newblock Phys. Rev. C 78 (2008) 044607.

\bibitem{montagnoli2009fusion}
G.~Montagnoli, S.~Beghini, B.~Guiot, et al.,
\newblock AIP Conf. Proc. 1098 (2009) 38.

\bibitem{PhysRevC.82.064609}
G.~Montagnoli, A.~M. Stefanini, L.~Corradi, et al.,
\newblock Phys. Rev. C 82 (2010) 064609.

\bibitem{PhysRevC.62.014601}
A.~M. Stefanini, L.~Corradi, A.~M. Vinodkumar, et al.,
\newblock Phys. Rev. C 62 (2000) 014601.

\bibitem{CAVALLARO1990174}
S.~Cavallaro, M.~Sperduto, B.~Delaunay, et al.,
\newblock Nucl. Phys. A 513 (1990) 174.

\bibitem{PhysRevC.14.1808}
W.~Scobel, H.~H. Gutbrod, M.~Blann, et al.,
\newblock Phys. Rev. C 14 (1976) 1808.

\bibitem{PhysRevC.41.2164}
E.~M. Szanto, R.~L. Neto, M.~C.~S. Figueira, et al.,
\newblock Phys. Rev. C 41 (1990) 2164.

\bibitem{PhysRevC.11.1701}
W.~Scobel, A.~Mignerey, M.~Blann, et al.,
\newblock Phys. Rev. C 11 (1975) 1701.

\bibitem{PhysRevC.102.024615}
R.~N. Sahoo, M.~Kaushik, A.~Sood, et al.,
\newblock Phys. Rev. C 102 (2020) 024615.

\bibitem{PhysRevC.63.054611}
E.~Mart\'{\i}nez-Quiroz, E.~F. Aguilera, J.~J. Kolata, et al.,
\newblock Phys. Rev. C 63 (2001) 054611.

\bibitem{mahon1997fusion}
J.~Mahon, L.~Lee~Jr, J.~Liang, et al.,
\newblock J. Phys. G: Nucl. Part. Phys. 23 (1997) 1215.

\bibitem{REISDORF1985212}
W.~Reisdorf, F.~Hessberger, K.~Hildenbrand, et al.,
\newblock Nucl. Phys. A 438 (1985) 212.

\bibitem{stokstad1980sub}
R.~Stokstad, W.~Reisdorf, K.~Hildenbrand, et al.,
\newblock Z. Phys. A 295 (1980) 269.

\bibitem{CLERC1984571}
H.-G. Clerc, J.~Keller, C.-C. Sahm, et al.,
\newblock Nucl. Phys. A 419 (1984) 571.

\bibitem{PhysRevC.30.1223}
H.~A. Aljuwair, R.~J. Ledoux, M.~Beckerman, et al.,
\newblock Phys. Rev. C 30 (1984) 1223.

\bibitem{PhysRevC.85.024607}
G.~Montagnoli, A.~M. Stefanini, C.~L. Jiang, et al.,
\newblock Phys. Rev. C 85 (2012) 024607.

\bibitem{PhysRevC.82.041601}
C.~L. Jiang, A.~M. Stefanini, H.~Esbensen, et al.,
\newblock Phys. Rev. C 82 (2010) 041601.

\bibitem{PhysRevC.65.011601}
M.~Trotta, A.~M. Stefanini, L.~Corradi, et al.,
\newblock Phys. Rev. C 65 (2001) 011601.

\bibitem{R_Vandenbosch_1997}
R.~Vandenbosch, A.~A. Sonzogni and J.~D. Bierman.
\newblock J. Phys. G: Nucl. Part. Phys. 23 (1997) 1303.

\bibitem{PhysRevC.20.2219}
B.~Sikora, J.~Bisplinghoff, W.~Scobel, et al.,
\newblock Phys. Rev. C 20 (1979) 2219.

\bibitem{PhysRevC.90.044610}
D.~Bourgin, S.~Courtin, F.~Haas, et al.,
\newblock Phys. Rev. C 90 (2014) 044610.

\bibitem{TIMMERS199735}
H.~Timmers, L.~Corradi, A.~Stefanini, et al.,
\newblock Phys. Lett. B 399 (1997) 35.

\bibitem{PhysRevC.76.014610}
A.~M. Stefanini, B.~R. Behera, S.~Beghini, et al.,
\newblock Phys. Rev. C 76 (2007) 014610.

\bibitem{STEFANINI2014639}
A.~Stefanini, G.~Montagnoli, H.~Esbensen, et al.,
\newblock Phys. Lett. B 728 (2014) 639.

\bibitem{A_M_Stefanini_1997}
A.~M. Stefanini.
\newblock J. Phys. G: Nucl. Part. Phys. 23 (1997) 1401.

\bibitem{SCARLASSARA200099}
F.~Scarlassara, S.~Beghini, G.~Montagnoli, et al.,
\newblock Nucl. Phys. A 672 (2000) 99.

\bibitem{PhysRevC.54.3068}
J.~D. Bierman, P.~Chan, J.~F. Liang, et al.,
\newblock Phys. Rev. C 54 (1996) 3068.

\bibitem{PhysRevC.45.2861}
A.~J. Pacheco, J.~O. Fern\'andez~Niello, D.~E. DiGregorio, et al.,
\newblock Phys. Rev. C 45 (1992) 2861.

\bibitem{PhysRevC.86.034608}
K.~Nishio, S.~Mitsuoka, I.~Nishinaka, et al.,
\newblock Phys. Rev. C 86 (2012) 034608.

\bibitem{STEFANINI200995}
A.~Stefanini, G.~Montagnoli, R.~Silvestri, et al.,
\newblock Phys. Lett. B 679 (2009) 95.

\bibitem{PhysRevC.73.034606}
A.~M. Stefanini, F.~Scarlassara, S.~Beghini, et al.,
\newblock Phys. Rev. C 73 (2006) 034606.

\bibitem{TROTTA2004245}
M.~Trotta, A.~Stefanini, L.~Corradi, et al.,
\newblock Nucl. Phys. A 734 (2004) 245.

\bibitem{stefanini2005fusion}
A.~Stefanini, M.~Trotta, B.~Behera, et al.,
\newblock Eur. Phys. J. A 23 (2005) 473.

\bibitem{PROKHOROVA200845}
E.~Prokhorova, A.~Bogachev, M.~Itkis, et al.,
\newblock Nucl. Phys. A 802 (2008) 45.

\bibitem{PhysRevC.65.034609}
A.~M. Stefanini, M.~Trotta, L.~Corradi, et al.,
\newblock Phys. Rev. C 65 (2002) 034609.

\bibitem{PhysRevC.92.064607}
A.~M. Stefanini, G.~Montagnoli, L.~Corradi, et al.,
\newblock Phys. Rev. C 92 (2015) 064607.

\bibitem{PhysRevC.53.803}
A.~M. Vinodkumar, K.~M. Varier, N.~V. S.~V. Prasad, et al.,
\newblock Phys. Rev. C 53 (1996) 803.

\bibitem{PhysRevC.81.037601}
A.~M. Stefanini, G.~Montagnoli, L.~Corradi, et al.,
\newblock Phys. Rev. C 81 (2010) 037601.

\bibitem{PhysRevC.23.1581}
M.~Beckerman, J.~Ball, H.~Enge, et al.,
\newblock Phys. Rev. C 23 (1981) 1581.

\bibitem{PhysRevLett.45.1472}
M.~Beckerman, M.~Salomaa, A.~Sperduto, et al.,
\newblock Phys. Rev. Lett. 45 (1980) 1472.

\bibitem{PhysRevC.25.837}
M.~Beckerman, M.~Salomaa, A.~Sperduto, et al.,
\newblock Phys. Rev. C 25 (1982) 837.

\bibitem{scarlassara1991fusion}
F.~Scarlassara, S.~Beghini, F.~Soramel, et al.,
\newblock Z. Phys. A 338 (1991) 171.

\bibitem{REHM199331}
K.~Rehm, H.~Esbensen, J.~Gehring, et al.,
\newblock Phys. Lett. B 317 (1993) 31.

\bibitem{PhysRevC.36.1379}
F.~L.~H. Wolfs,
\newblock Phys. Rev. C 36 (1987) 1379.

\bibitem{PhysRevC.34.2155}
K.~T. Lesko, W.~Henning, K.~E. Rehm, et al.,
\newblock Phys. Rev. C 34 (1986) 2155.

\bibitem{PhysRevC.91.044602}
C.~L. Jiang, A.~M. Stefanini, H.~Esbensen, et al.,
\newblock Phys. Rev. C 91 (2015) 044602.

\bibitem{PhysRevLett.89.052701}
C.~L. Jiang, H.~Esbensen, K.~E. Rehm, et al.,
\newblock Phys. Rev. Lett. 89 (2002) 052701.

\bibitem{stefanini2013fusion}
A.~Stefanini, G.~Montagnoli, F.~Scarlassara, et al.,
\newblock Eur. Phys. J. A 49 (2013) 1.

\bibitem{PhysRevLett.93.012701}
C.~L. Jiang, K.~E. Rehm, R.~V.~F. Janssens, et al.,
\newblock Phys. Rev. Lett. 93 (2004) 012701.

\bibitem{STEFANINI1992453}
A.~Stefanini, L.~Corradi, D.~Ackermann, et al.,
\newblock Nucl. Phys. A 548 (1992) 453.

\bibitem{JANSSENS198616}
R.~Janssens, R.~Holzmann, W.~Henning, et al.,
\newblock Phys. Lett. B 181 (1986) 16.

\bibitem{PhysRevLett.50.1563}
W.~S. Freeman, H.~Ernst, D.~F. Geesaman, et al.,
\newblock Phys. Rev. Lett. 50 (1983) 1563.

\bibitem{PhysRevLett.55.803}
K.~T. Lesko, W.~Henning, K.~E. Rehm, et al.,
\newblock Phys. Rev. Lett. 55 (1985) 803.

\bibitem{PhysRevC.28.1963}
M.~Beckerman, M.~K. Salomaa, J.~Wiggins, et al.,
\newblock Phys. Rev. C 28 (1983) 1963.

\bibitem{PhysRevC.29.1938}
M.~Beckerman, J.~Wiggins, H.~Aljuwair, et al.,
\newblock Phys. Rev. C 29 (1984) 1938.

\bibitem{REISDORF1985154}
W.~Reisdorf, F.~Hessberger, K.~Hildenbrand, et al.,
\newblock Nucl. Phys. A 444 (1985) 154.

\bibitem{PhysRevC.85.054603}
J.~J. Kolata, A.~Roberts, A.~M. Howard, et al.,
\newblock Phys. Rev. C 85 (2012) 054603.

\bibitem{PhysRevLett.107.202701}
Z.~Kohley, J.~F. Liang, D.~Shapira, et al.,
\newblock Phys. Rev. Lett. 107 (2011) 202701.

\bibitem{liang2004enhanced}
J.~Liang, D.~Shapira, C.~Gross, et al.,
\newblock Nucl. Phys. A  (2004) 103.

\bibitem{PhysRevC.75.054607}
J.~F. Liang, D.~Shapira, J.~R. Beene, et al.,
\newblock Phys. Rev. C 75 (2007) 054607.

\bibitem{PhysRevC.87.064612}
Z.~Kohley, J.~F. Liang, D.~Shapira, et al.,
\newblock Phys. Rev. C 87 (2013) 064612.

\bibitem{shapira2005measurement}
D.~Shapira, J.~Liang, C.~Gross, et al.,
\newblock Eur. Phys. J. A 25 (2005) 241.

\bibitem{BOCK1982334}
R.~Bock, Y.~Chu, M.~Dakowski, et al.,
\newblock Nucl. Phys. A 388 (1982) 334.

\end{thebibliography}
\end{document}